\documentclass{elsart4-1}

\usepackage{graphicx}
\usepackage{epsfig}

\usepackage{amssymb}

\usepackage[english,francais]{babel}

\graphicspath{{./figs/}}
\usepackage{xspace}

\newcommand{\Swift}{\emph{Swift}\xspace}
\newcommand{\Fermi}{\emph{Fermi}\xspace}
\newcommand{\Nustar}{\emph{NuSTAR}\xspace}
\newcommand{\eiso}{E_\mathrm{iso}}
\newcommand{\alex}{\alpha_\mathrm{EX}}
\newcommand{\alt}{\delta}
\newcommand{\all}{\delta_\infty}

\newtheorem{e-proposition}[theorem]{Proposition}

\newtheorem{e-definition}[theorem]{Definition\rm}

\renewcommand{\theequation}{\arabic{equation}}
\def\og{\leavevmode\raise.3ex\hbox{$\scriptscriptstyle\langle\!\langle$~}}
\def\fg{\leavevmode\raise.3ex\hbox{~$\!\scriptscriptstyle\,\rangle\!\rangle$}}

\begin{document}

\centerline{Astrophysics}
\begin{frontmatter}



\selectlanguage{english}
\title{Gamma-Ray Bursts at high and very high energies}


\selectlanguage{english}
\author{Fr\'ed\'eric Piron}
\ead{piron@in2p3.fr}
\address{LUPM, CC 72, CNRS/IN2P3, Universit\'e de Montpellier, place E. Bataillon\\
34095 Montpellier CEDEX 05, France}

\medskip
\begin{center}
{\small Received *****; accepted after revision +++++}
\end{center}

\begin{abstract}
Gamma-Ray Bursts (GRBs) are extra-galactic and extremely energetic transient emissions of gamma rays, which are thought to be
associated with the death of massive stars or the merger of compact objects in binary systems.
Their huge luminosities involve the presence a newborn stellar-mass black hole emitting a relativistic collimated outflow, which
accelerates particles and produces non-thermal emissions from the radio domain to the highest energies.
In this article, I review recent progresses in the understanding of GRB jet physics above 100\,MeV, based on \Fermi observations
of bright GRBs.
I discuss the physical implications of these observations and their impact on GRB modeling, and I present some prospects for GRB
observation at very high energies in the near future.

{\it To cite this article: F. Piron, C. R. Physique XX (2015).}

\vskip 0.5\baselineskip

\selectlanguage{francais}
\noindent{\bf R\'esum\'e}
\vskip 0.5\baselineskip
\noindent
{\bf Les sursauts gamma \`a haute et tr\`es haute \'energie.}
Les sursauts gamma sont des ph\'enom\`enes explosifs extr\^emement \'energ\'etiques et des \'emissions transitoires
de rayonnement gamma d'origine extra-galactique.
Ils sont associ\'es \`a la fin de vie d'\'etoiles massives ou \`a la fusion d'objets compacts d'un m\^eme syst\`eme binaire.
Leur grande luminosit\'e implique la pr\'esence d'un trou noir de masse stellaire nouvellement form\'e \'emettant un \'ecoulement
relativiste collimat\'e qui acc\'el\`ere les particules et produit une \'emission non thermique du domaine radio jusqu'aux plus
hautes \'energies.
Dans cet article, je passe en revue les progr\`es r\'ecents dans la compr\'ehension de la physique des jets de sursauts gamma
au-dessus de 100\,MeV, sur la base d'observations de sursauts brillants avec \Fermi.
Je discute les implications physiques imm\'ediates de ces observations et leurs cons\'equences sur la mod\'elisation, et je
pr\'esente les perspectives pour l'observation de sursauts gamma aux tr\`es hautes \'energies dans les ann\'ees \`a venir.

{\it Pour citer cet article~: F. Piron, C. R. Physique XX (2015).}

\vskip 0.5\baselineskip
\noindent{\small{\it Keywords~:} gamma-ray astronomy; gamma-ray bursts; relativistic jets}
\vskip 0.5\baselineskip
\noindent{\small{\it Mots-cl\'es~:} astronomie gamma; sursauts gamma; jets relativistes}

\end{abstract}
\end{frontmatter}


\selectlanguage{english}

\section{Introduction}
\label{sec:intro}
\begin{figure}[!t]
  \centering
  \includegraphics[width=0.7\linewidth]{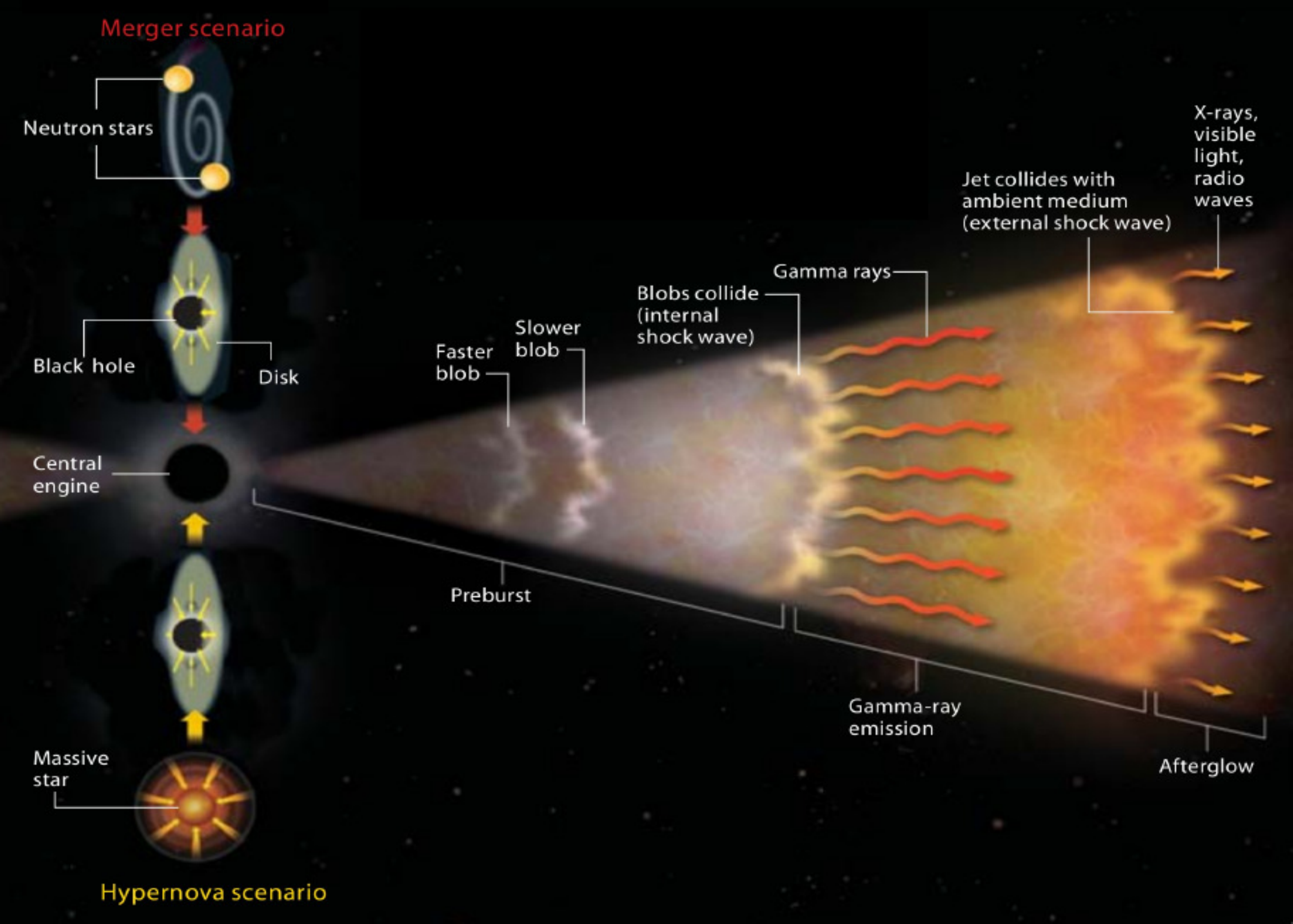}
  \caption[]{Illustration of the internal / external shock scenario for gamma-ray bursts (credit NASA).}
  \label{fig:paradigm}
\end{figure}

Gamma-Ray Bursts (GRBs) are powerful high-energy transient emissions from sources at cosmological distances.
They appear randomly on the celestial sphere, and they are characterized by a short phase of intense and erratic emission in hard
X rays and gamma rays, lasting from few milliseconds to hundreds of seconds.
Following this so-called prompt phase, GRBs exhibit a long-lasting activity (the afterglow phase) where the observed flux
decreases rapidly in time, with an emission peak energy shifting to longer wavelengths (X rays, visible and radio) on time scales
spanning from hours to weeks.\\

Assuming that GRBs radiate isotropically, their energy release in X rays and gamma rays (typically $\sim$$10^{44-47}$\,J $=10^{51-54}$\,erg) exceeds hundred
times the total energy radiated by a supernova ($\sim$$10^{42}$\,J $=10^{49}$\,erg).
The widely accepted scenario~\cite{piran2004,nakar2007,vedrenne2009} which is invoked to explain GRB huge luminosities involves a newborn
stellar-mass black hole emitting a relativistic collimated outflow (figure~\ref{fig:paradigm}).
The GRB progenitors as well as the physical conditions which are required to produce and accelerate the relativistic jet are
nevertheless still unclear.
The distribution of GRB duration was found to be bimodal~\cite{kouveliotou1993,avk2014}, revealing the existence of two distinct
populations.
Long GRBs (i.e., with typical prompt emission durations $\gtrsim2$\,s) are believed to be produced by the collapse of fast
rotating massive stars ($\gtrsim30\,M_\odot$, of Wolf-Rayet type), as suggested by their association with nearby core-collapsed
supernovae of types Ib/Ic.
Short GRB progenitors are more elusive and may be connected to the merging of two compact objects in binary systems (two neutron
stars or a neutron star and a stellar-mass black hole).
The new generation of gravitational wave detectors such as Advanced Virgo /
LIGO\footnote{\tt{https://wwwcascina.virgo.infn.it/advirgo},~\tt{https://www.advancedligo.mit.edu}} should be able to detect the
corresponding emission from short GRBs, and to help shed light on their progenitors in the coming years~\cite{gwsearch}.\\

The highly variable gamma-ray emission characteristic of the GRB prompt phase is associated with the dissipation of the
jet internal energy in midly relativistic shocks taking place at a distance $R\approx10^{12-13}$\,m from the central engine,
where particles are accelerated and emit non-thermal radiations.
The properties of the jet (speed, structure, collimation, composition, energy reservoirs, magnetization, emission sites) as well as the
micro-physics (energy dissipation mechanisms, shock acceleration efficiency, radiation processes and internal opacity effects, role
of magnetic fields, etc) are not precisely known.
The afterglow phase is associated with the jet deceleration at a distance $R\approx10^{14-15}$\,m from the central engine, where an
external shock is formed by the interaction of the jet with the circum-burst medium.
This relativistic forward shock accelerates electrons which radiate synchrotron emission observed from the radio to the gamma-ray
domain.\\
\begin{figure}[!t]
  \centering
  \includegraphics[width=0.58\linewidth]{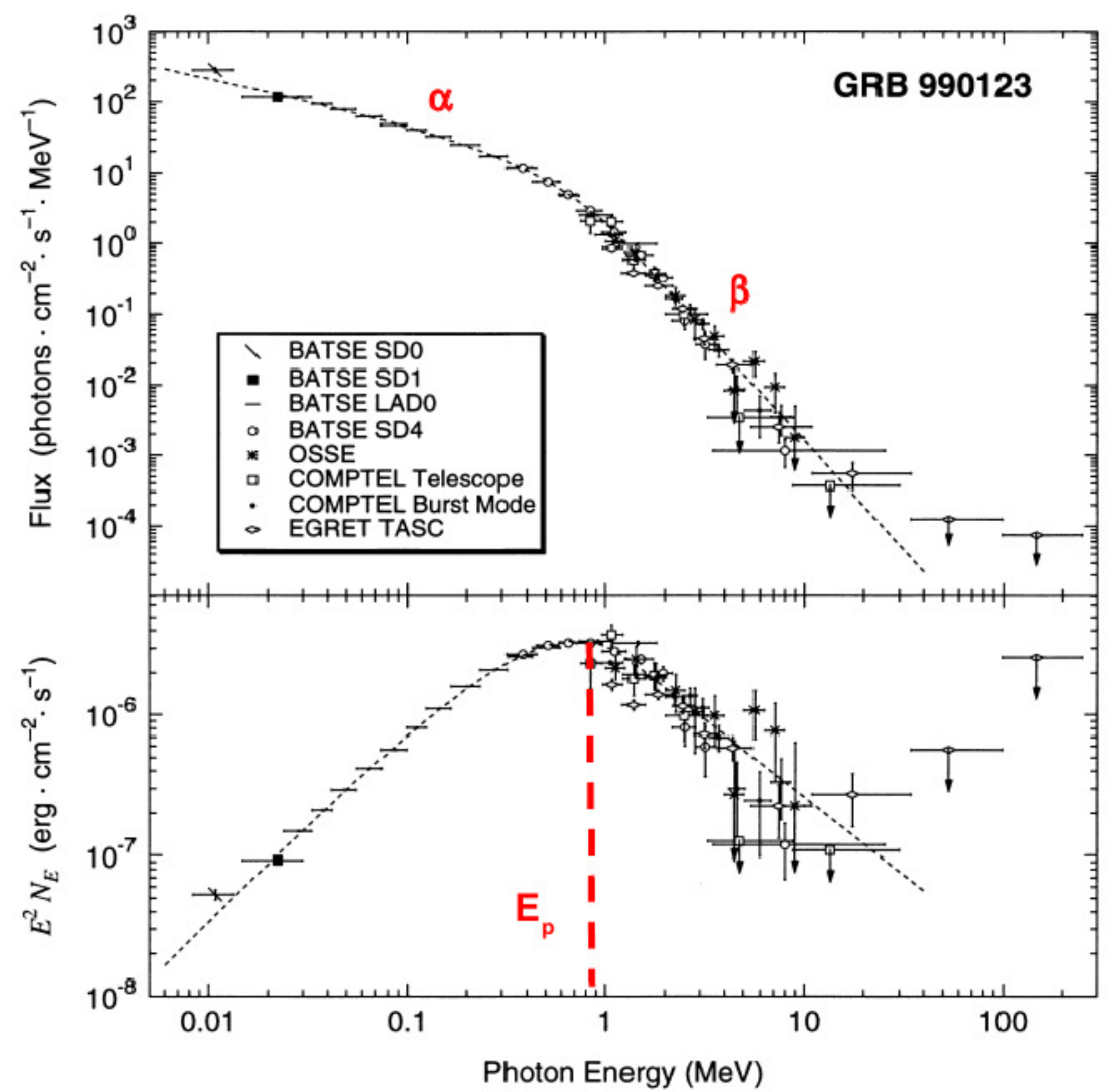}
  \includegraphics[width=0.4\linewidth]{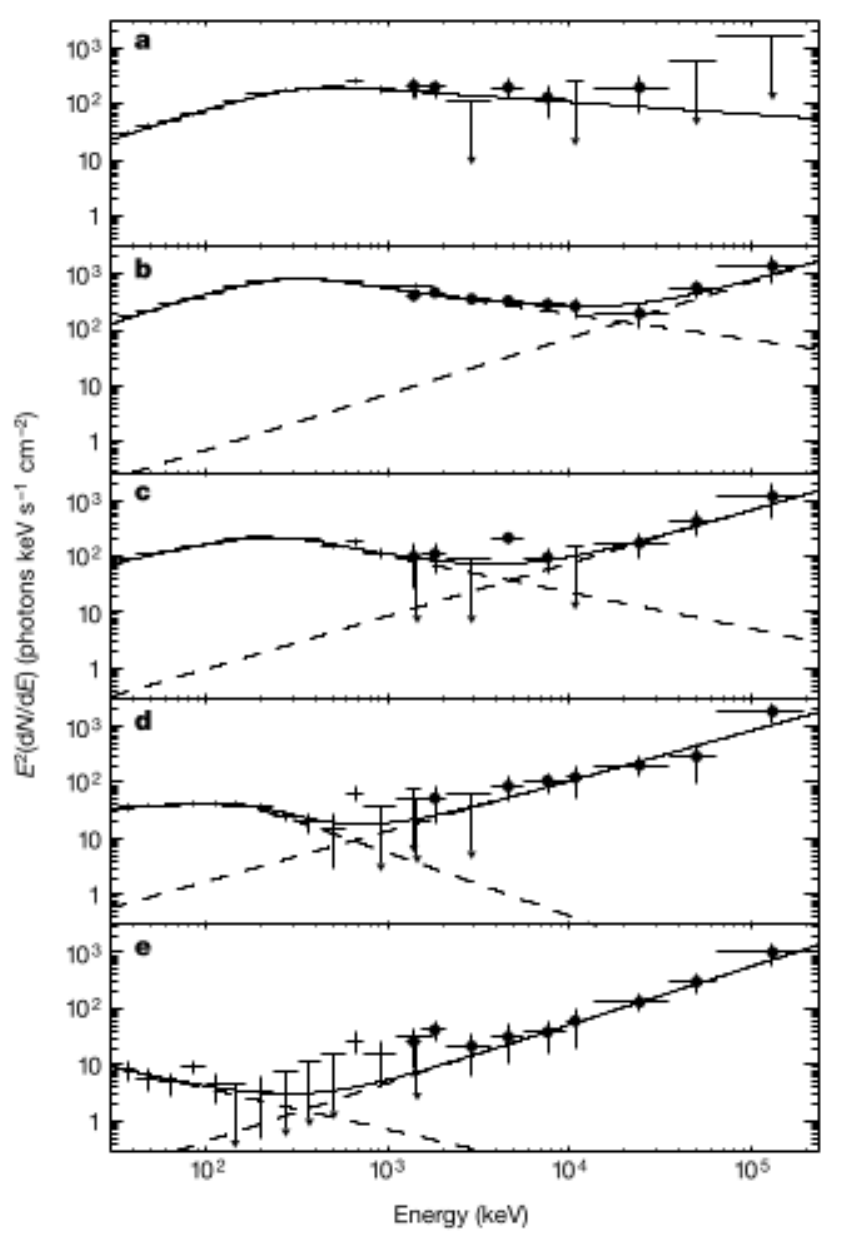}
  \caption[]{
    Left~: GRB\,990123 photon spectrum (top panel) and spectral energy distribution (bottom panel, in
    erg\,cm$^{-2}$\,s$^{-1}=10^{-3}$\,W\,m$^{-2}$) as measured by the four instruments
    onboard the Compton Gamma-Ray Observatory~\cite{briggs1999}.
    Right~: GRB\,941017 spectral energy distributions as measured by the CGRO/BATSE and the CGRO/EGRET between $-$18\,s and
    $+$211\,s (from top to bottom)~\cite{gonzalez03}.
  }
  \label{fig:990123+941017}
\end{figure}

The study of acceleration and emission processes in GRB jets requires well-sampled (both spectrally and temporally)
multi-wavelength observations of objects with measured redshifts.
Significant progress has been made possible thanks to the space missions \Fermi (launched in 2008) and \Swift (launched in
2004)\footnote{Detailed reviews can be found elsewhere, e.g., in~\cite{kumar2014}.}.
Following a previous ``Comptes Rendus Physique'' on \Fermi results~\cite{cras2011}, this article focuses on GRB physics\footnote{The
  detection of high-energy photons from distant GRBs is also a powerful tool for non-GRB science.
  Specifically, they have been used to probe the quantum-gravitational nature of space-time and
  to test the existence of Lorentz Invariance Violation (LIV) as a consequence of some Quantum Gravity theories.
  In addition, gamma rays with energies above $\sim$10\,GeV can be absorbed by the Extra-galactic Background Light (EBL) when
  travelling from the emitting region to the observer, providing useful constraints on this cosmic diffuse radiation field, which
  results from the emission of the first stars and its subsequent reprocessing by dust in the inter-stellar medium.
  Both topics (LIV and EBL) are addressed in detail in another article in this volume~\cite{cras2015-probes}.}
above 100\,MeV (``high energies'' hereafter) up to the very high-energy gamma-ray domain (above 100\,GeV).
In section~\ref{sec:egret}, I briefly review some of the few GRB detections at high energies before the \Fermi era, and the
questions they raised.
\Fermi observations of GRBs have greatly improved the detection statistics, and they constitute the best set of high-energy data so far.
In section~\ref{sec:fermi}, I recall the main findings from this mission, including notable GRBs as well as population
studies, and I present the common properties in GRB temporal and spectral behavior at high energies.
Section~\ref{sec:grbphys} addresses the physical implications of these observations and their impact on GRB modeling.
In section~\ref{sec:vhe}, I discuss the prospects for GRB observation at very high energies with the new generation of ground-based
Cherenkov telescopes in the coming years, before concluding in section~\ref{sec:concl}.

\section{GRBs at high energies before the \Fermi era}
\label{sec:egret}
Before the advent of \Fermi, non-thermal spectra in the prompt phase were usually
represented in the keV--MeV energy range by the phenomenological Band function~\cite{band93}, which is composed of two
smoothly-connected power laws with respective indices $\alpha$ and $\beta$ (figure~\ref{fig:990123+941017}, left):
\begin{equation}
\label{eq:band}
\frac{dN}{dE}\big(E| N_0,E_p,\alpha,\beta\big) = N_0 \left\{
\begin{array}{l l}
   E^{\alpha} \exp[-E(2+\alpha)/E_p], &  E \le E_p {{\alpha-\beta}\over {2+\alpha}} \\
\\
   E^\beta \left[E_p {{\alpha-\beta}\over {2+\alpha}}\right]^{(\alpha-\beta)}
   \exp[\beta-\alpha] , & E > E_p {{\alpha-\beta}\over {2+\alpha}}
\end{array}
\right.
\end{equation}
where $\displaystyle\frac{dN}{dE}$ is the photon spectrum\footnote{In this article, photon indices will be always chosen negative,
  i.e. $\displaystyle\frac{dN}{dE}\propto E^\gamma$ with $\gamma<0$, following the definition of the Band function.} (in m$^{-2}$\,s$^{-1}$\,keV$^{-1}$)
and $E_p$ is the peak energy of the spectral energy distribution $\displaystyle E^2\frac{dN}{dE}$, ranging from a few hundreds of keV to a few MeV.
At higher energies, GRB emission had been detected in a few distinct cases with the Energetic Gamma-Ray Experiment Telescope
(EGRET, 30\,MeV--30\,GeV) onboard the Compton Gamma-Ray Observatory (CGRO, 1991--2000)\footnote{And in one case (GRB\,080514B) by the
  GRID instrument onboard Astro-rivelatore Gamma a Immagini LEggero (AGILE)~\cite{giuliani08}.}.
The high-energy emission from GRB\,930131~\cite{sommer94} was consistent with an extrapolation from its keV--MeV spectrum as
measured by the CGRO's Burst And Transient Source Experiment (BATSE).
In the case of GRB\,941017~\cite{gonzalez03}, an additional and hard power law was observed up to 200\,MeV without any spectral
attenuation (figure~\ref{fig:990123+941017}, right).
This extra component had a much larger energetics and lasted significantly longer ($\sim$200\,s) than the Band component
detected by the CGRO/BATSE at sub-MeV energies.
Moreover, both components appeared uncorrelated, the extra component exhibiting a temporal stability that contrasts with the
gradual decrease of the Band component intensity and peak energy.
Since the Band component is commonly attributed to the synchrotron emission of internal-shock accelerated electrons (see
section~\ref{sec:grbint}), these observations ruled out a possible inverse Compton origin of the extra component, as is for
instance expected from simple Synchrotron Self Compton (SSC) models.
Alternate interpretations were considered, such as a possible hadronic origin (e.g., gamma-ray emission of secondary particles
produced in internal cascades initiated by accelerated protons or ions) or an external shock emission~\cite{granot2003}.
In the case of GRB\,940217~\cite{hurley94}, a delayed high-energy emission was observed up to $\sim$90 minutes, with an 18\,GeV
photon detected after $\sim$75 minutes. 
Possible interpretations of these observations included a GRB late activity, a hadronic spectral component, or an SSC emission in
the afterglow phase.\\

These pioneering studies illustrate the diversity in GRB spectral and temporal properties at high energies, which can differ
from the main emission spectral component that culminates at a few hundreds of keV.
They opened several questions about GRB jet physics, anticipating the advances which were later provided by \Fermi:\\
\begin{itemize}
\item What is the nature of the accelerated particles responsible for the high-energy emissions?\\
\item Where and how are these particles accelerated? Within internal or external shocks? What is their spectrum?\\
\item When do internal shocks end (prompt phase) and external shocks start (afterglow phase)?\\
\item What are the dominant radiative processes at high energies? At which distance / radius from the central engine are non-thermal
  emissions produced?\\
\item Is the extra spectral component common to most GRBs?\\
\item What is the maximum detectable energy and what is the shape of the spectral attenuation at the highest energies?
  Does it mark the end of the particle spectrum, is it caused by a lower radiation efficiency or by opacity effects within the source?\\
\end{itemize}

Precious information on the gamma-ray opacity to pair production ($\gamma+\gamma\rightarrow e^++e^-$) is provided by the
detection of high-energy photons, which imply strong evidence for relativistic outflows as the sites of GRB prompt emission.
Together with GRB variability and brightness at high energies, these photons can be used to set strong limits on the jet
bulk Lorentz factor $\Gamma$.
For a source at rest, the short variability time scale $t_v$ observed in the prompt light curves gives a limit on the size of the
emitting zone $R<c\,t_{v}$, based on a simple causality argument.
Combined with the large luminosities $L_\mathrm{iso}\sim10^{43-46}$\,W ($10^{50-53}$\,erg\,s$^{-1}$) inferred by assuming an isotropic emission, this
compactness would be sufficient for gamma rays of energy $E$ to annihilate in pairs with dense fields of softer
photons. Since pair production is most efficient for soft target photons of energies $\epsilon\simeq m_e^2c^4/E$, the implied
optical depth is:
\begin{eqnarray}
\label{eq:opacity}
  \tau_{\gamma\gamma}(E) & \simeq & \sigma_\mathrm{T}\,n_\mathrm{ph}(\epsilon)\,R\nonumber\\
                             & = & \sigma_\mathrm{T}\,\frac{L_{\mathrm{iso},{\epsilon}}}{4\pi\,m_e\,c^3\,R}\nonumber\\
                             & > &
  10^{13}\left(\frac{L_{\mathrm{iso},{\epsilon}}}{10^{44}\,\mathrm{W}}\right)\left(\frac{t_v}{10\,\mathrm{ms}}\right)^{-1}
\end{eqnarray}
where $n_\mathrm{ph}(\epsilon)$ is the soft photon number density at energy $\epsilon$ and where the pair production cross section
has been approximated as the Thomson cross section $\sigma_\mathrm{T}=0.665$\,barns ($1$\,barn $=10^{-28}$\,m$^2$).
The opacity in equation~\ref{eq:opacity} is huge and would produce a thermal spectrum, in contradiction with the non-thermal power-law spectra observed up to high energies.
Considering a source moving at relativistic speed towards the observer can solve this well-known compactness problem.
In this case and, e.g., for a pure Band spectrum\footnote{See the Supporting Online Material in~\cite{080916C} for a detailed
  computation.}, the opacity is reduced by a factor $\displaystyle\Gamma^{2(1-\beta)}$, and it can be less than unity for a
typical index $\beta\simeq-$2.3 combined with a minimum value of the jet Lorentz factor $\Gamma>\Gamma_\mathrm{min}$$\sim$100
(increasing with $E$, $1/t_v$, the redshift and the source intensity).

\section{GRB observations with \Fermi}
\label{sec:fermi}
\subsection{The GBM and LAT instruments}
\label{sec:grbinst}
\begin{figure}[!t]
  \centering
  \includegraphics[width=.45\linewidth]{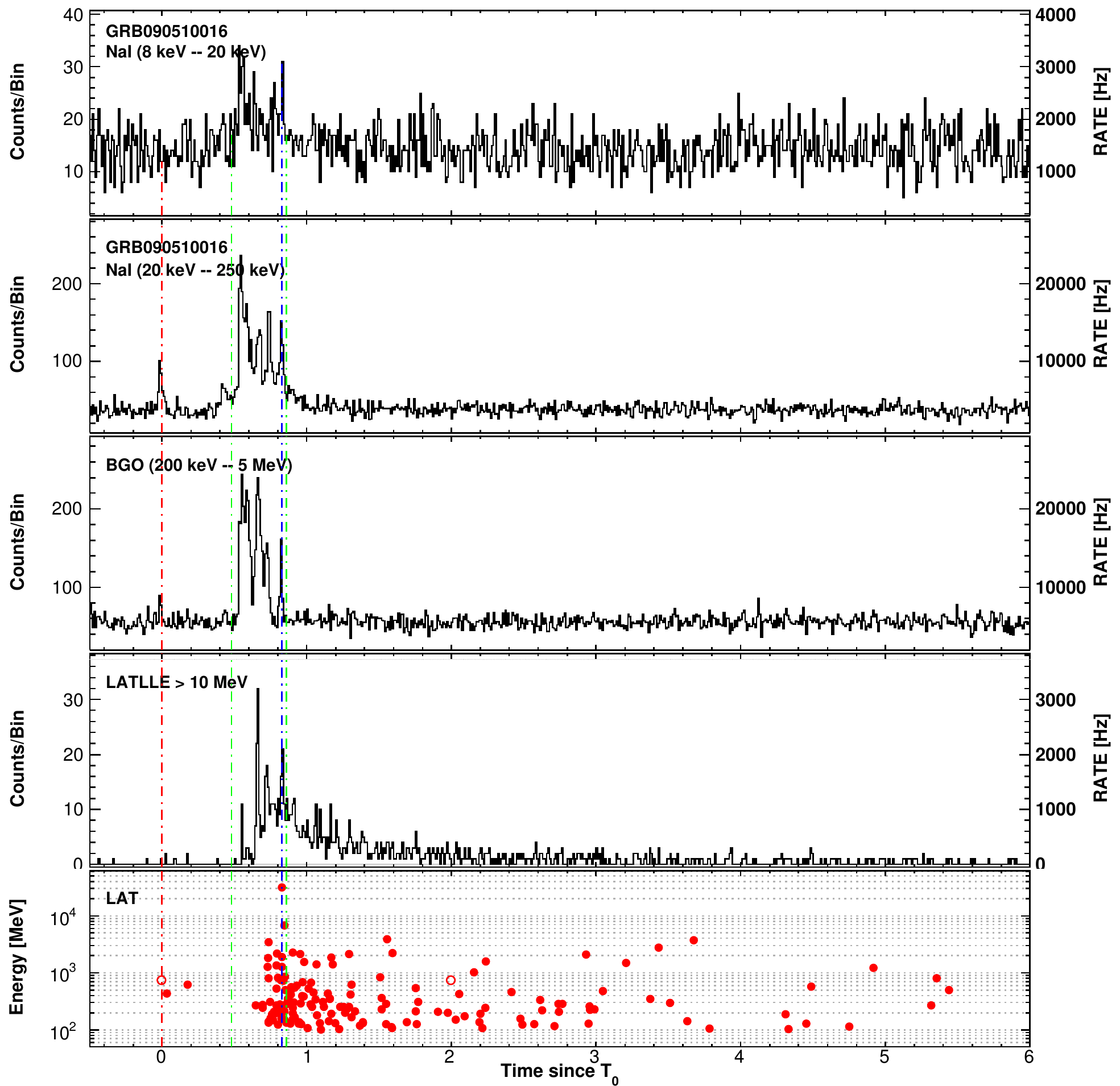}
  \includegraphics[width=.53\linewidth]{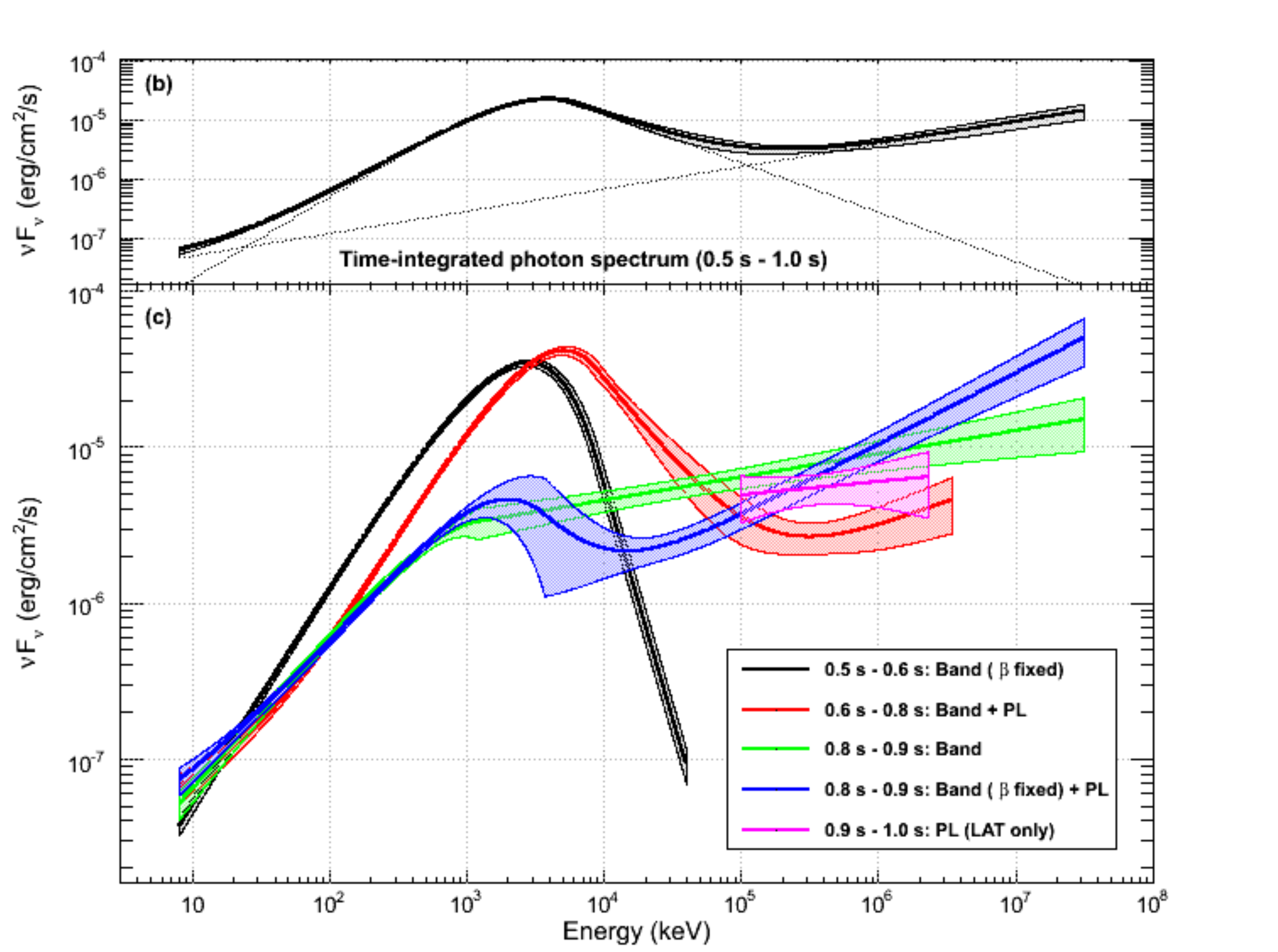}
  \caption{
    Left~: GRB\,090510 light curves as measured by the \Fermi/GBM and the \Fermi/LAT~\cite{grbcat}, from lowest to highest
    energies: sum of the counts in the GBM NaI detectors (first two panels), in the GBM BGO detector facing the burst (third
    panel), and in the LAT using all events passing the onboard filter for gamma rays (fourth panel).
    The last panel displays the energies of the individual photons which are well-reconstructed by the LAT above 100\,MeV.
    Right~: GRB\,090510 spectral energy distribution (in erg\,cm$^{-2}$\,s$^{-1}=10^{-3}$\,W\,m$^{-2}$) as
    measured by \Fermi~\cite{090510_prompt}. Each spectrum is represented by a 68\%
    confidence level contour corresponding to the best fitted spectral shape (i.e., a Band function with an extra power law
    at high energies).
  }
  \label{fig:090510}
\end{figure}
Launched in 2008, the \Fermi observatory is composed of two instruments which together cover more than 7 decades in energy.
The Gamma-ray Burst Monitor (GBM; \cite{meegan09}) is comprised of 14 scintillation detectors which monitor the entire sky that is
not occulted by the Earth.
Spectroscopy with the GBM makes use of 12 NaI detectors between 8\,keV and 1\,MeV, and of two BGO scintillators which are
sensitive to photons of energies between 150\,keV and 40\,MeV.
As a result, the GBM can measure spectra with high time resolution over nearly 5 decades in energy, covering the low energy domain
where most of the GRB emission takes place~\cite{goldstein2012,gruber2014}.
At higher energies, the Large Area Telescope (LAT; \cite{lat1,cras2015-space}) is a pair-conversion detector sensitive to gamma
rays of energies ranging from 20\,MeV to more than 300\,GeV.
The LAT broad energy range, large effective area ($\sim$0.9\,m$^2$ at peak), low deadtime per event ($\sim$27\,$\mu$s), wide
field of view ($\sim$2.4\,sr at 1\,GeV) and good angular resolution ($\sim$0.2$^\circ$ at 10\,GeV) are vastly improved in comparison
with those of its predecessor CGRO/EGRET. They provide more GRB detections and more photons detected from each GRB.
Moreover, \Fermi was designed with the capability to repoint in the direction of any bright GRB in order to keep its position near the
center of the LAT field of view and to observe its afterglow emission during several hours.\\

The GBM detects 240 bursts per year in average, including 45 short bursts~\cite{paciesas2012,avk2014}.
About half of these GRBs occur in the LAT field of view, $\sim$10\% of them being detected above 100\,MeV.
After 6.5 years of operations, the LAT has detected nearly 90
bursts\footnote{\tt{http://fermi.gsfc.nasa.gov/ssc/observations/types/grbs/lat\_grbs/table.php}.}, including 7 short bursts.
All LAT bright GRBs have benefited from accurate follow-up localizations by the narrow-field instruments onboard
\Swift~\cite{gehrels2004}, facilitating their distance measurement by ground-based optical telescopes. As of today, LAT GRB
redshifts range from $z=0.145$ (GRB\,130702A) to $z=4.35$ (GRB\,080916C).

\subsection{Some remarkable bursts}
\label{sec:grbind}
\begin{figure}[!t]
  \centering
  \includegraphics[width=.45\linewidth]{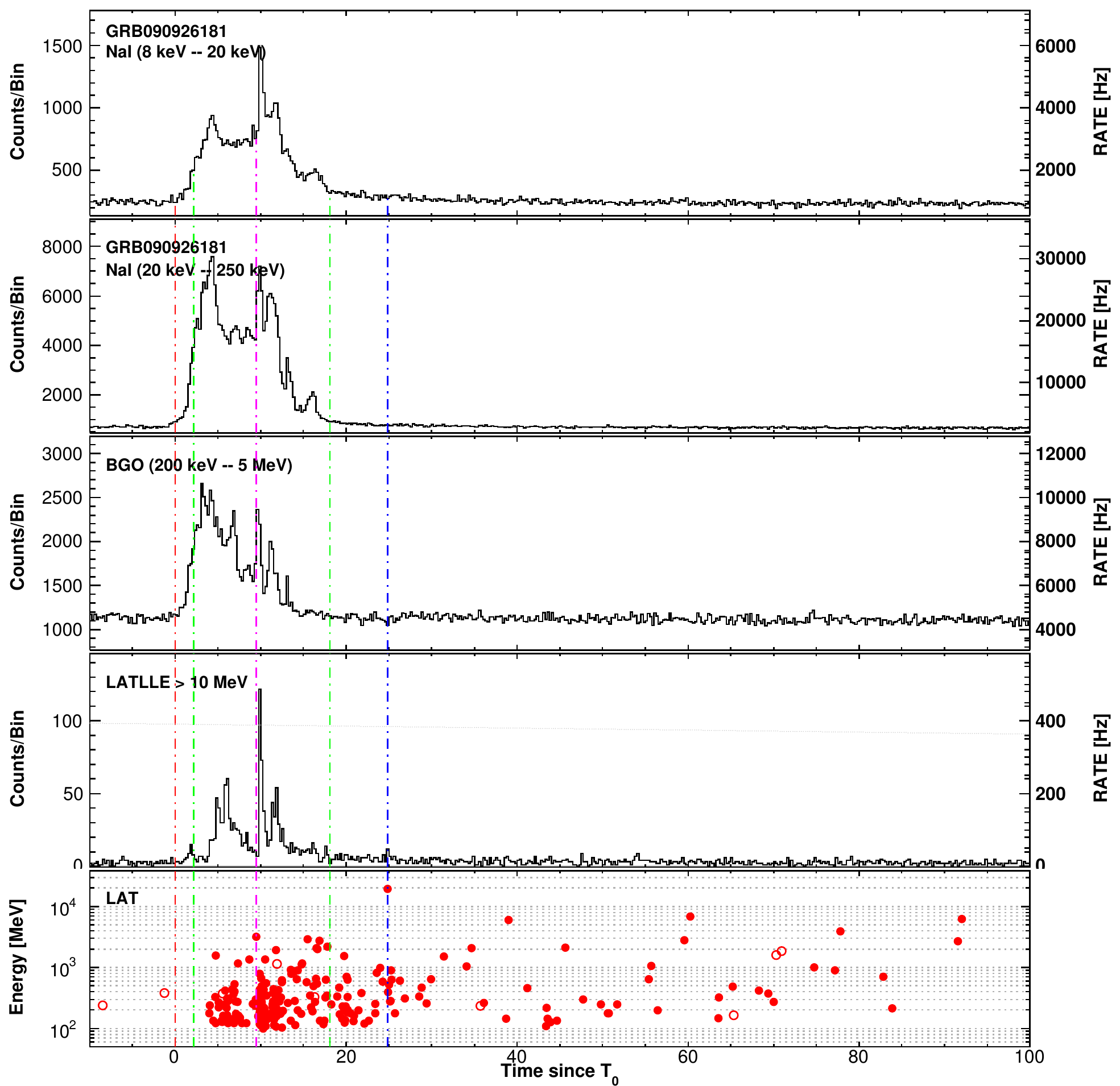}
  \includegraphics[width=.53\linewidth]{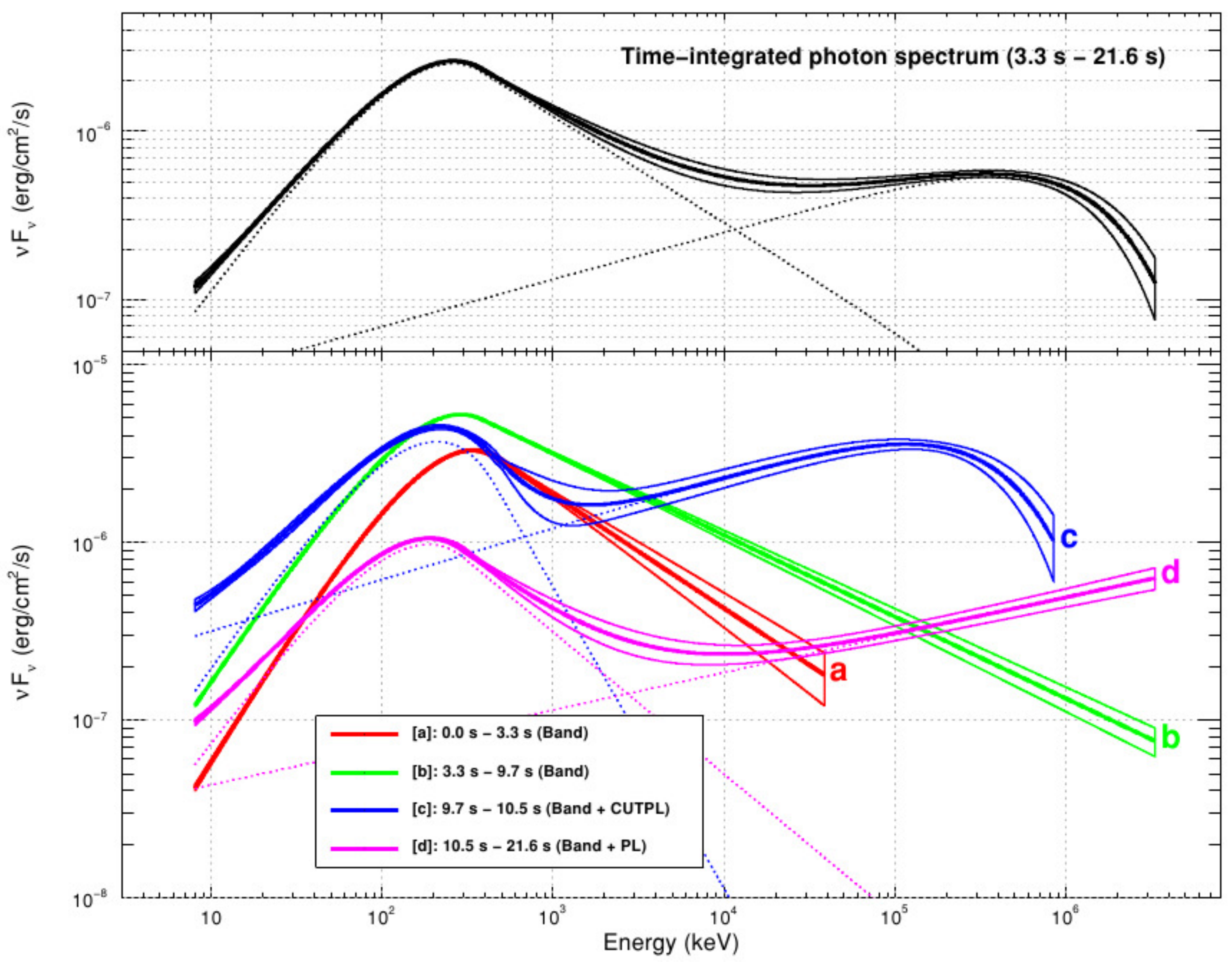}
  \caption{
    Left~: GRB\,090926A light curves as measured by the \Fermi/GBM and the \Fermi/LAT~\cite{grbcat} (see caption of
    figure~\ref{fig:090510}, left).
    Right~: GRB\,090926A spectral energy distribution (in erg\,cm$^{-2}$\,s$^{-1}=10^{-3}$\,W\,m$^{-2}$) as
    measured by \Fermi~\cite{090926A}. Each spectrum is represented by a 68\%
    confidence level contour corresponding to the best fitted spectral shape (i.e., a Band function with an extra power law
    at high energies, possibly including a spectral cutoff).
  }
  \label{fig:090926a}
\end{figure}
The long and bright GRB\,080916C is the second LAT-detected burst~\cite{080916C}.
Its high-energy emission extended up to 13.2\,GeV in the prompt phase, which implied the largest jet Lorentz factor ever measured,
$\Gamma_\mathrm{min}=870$.
GRB\,080916C is also the record holder in terms of energetics, with a source frame energy release\footnote{The isotropic
  equivalent energy is defined as~\cite{eiso}
  $\displaystyle\eiso=\frac{4\pi\,D_l^2}{1+z}\int_{E'_1/(1+z)}^{E'_2/(1+z)}E\,S(E)\,dE$, where $D_l$ is the luminosity distance,
  $S(E)$ is the time-integrated photon spectrum (in m$^{-2}$\,keV$^{-1}$) in the observer frame, and $[E'_1, E'_2]$ is the energy interval in the source frame.}
$\eiso=8.8\times10^{47}$\,J ($8.8\times10^{54}$\,erg) in the 10\,keV--10\,GeV energy band, which is equivalent to 4.9 times the Sun rest mass energy.
GRB\,080916C prompt emission spectrum features a faint extra power-law component~\cite{grbcat}.
Its high-energy emission is delayed by a few seconds with respect to its keV--MeV emission and it has been detected up to
1400\,s post-trigger, i.e. well after the GBM-detected emission has faded.
Subsequent GRB detections with the LAT have revealed that the last two characteristics are common to the vast majority of GRBs at
high energies (see section~\ref{sec:grbprop}).\\

GRB\,090510 is the first short and bright burst detected by the LAT~\cite{090510_prompt}, with an observed emission extending up to
31.3\,GeV in the prompt phase.
Its high-energy emission is delayed and temporally extended with respect to its keV--MeV emission (figure~\ref{fig:090510}, left).
Similarly to all LAT long and bright bursts, its prompt emission spectrum features an extra power-law component.
This spectral component dominates the Band function not only at high energies but also below $\sim$20\,keV
(figure~\ref{fig:090510}, right).
The afterglow emission of GRB\,090510 has been detected by the LAT up to 200\,s post-trigger and has been observed simultaneously by
\Swift. A forward shock synchrotron emission model (see section~\ref{sec:grbint}) has been successfully applied to these
multi-wavelength observations (from the visible domain to GeV energies)~\cite{090510_afterglow}.\\

The high-energy properties of the long and bright GRBs\,090902B~\cite{090902B} and 090926A~\cite{090926A} are very similar to
those of the two bursts discussed above: presence of an extra power-law component and delayed onset of the high-energy emission,
which persisted during 1\,ks and 5\,ks, respectively.
Both temporally extended emissions include multi-GeV photons, e.g., the most energetic event (33.4\,GeV) from GRBs\,090902B was
detected by the LAT 82\,s after the GBM trigger.
The prompt emission spectrum of GRB\,090902B is also peculiar since it contains an important flux in excess to the Band
function below $\sim$50\,keV, the origin of which is poorly understood.
Morever, the narrow spectral shape of GRB\,090902B around its peak energy $E_p$ is difficult to account for in the framework of internal
shock synchrotron models (see section~\ref{sec:grbint}).
GRB\,090926A is also remarkable, especially for the extreme variability of its prompt light curve.
A sharp pulse lasting less than 1\,s was observed $\sim$10\,s post-trigger across the whole spectrum (figure~\ref{fig:090926a}, left).
The temporal correlation (within less than 50\,ms) between the keV--MeV and GeV prompt emissions during this sharp
pulse, which coincides with the emergence of the extra power-law component (figure~\ref{fig:090926a}, right), suggests that
an important fraction of the high-energy emission has an internal shock origin in this case.
Morevover, GRB\,090926A is unique due to the attenuation of its prompt emission spectrum.
The spectral cutoff observed during the sharp pulse (in blue in figure~\ref{fig:090926a}, right)
provided a direct estimate of the jet Lorentz factor $\Gamma$ (i.e., not just a lower limit as discussed in section~\ref{sec:egret}),
ranging from 200 to 700 depending on the $\gamma\gamma$ absorption model which was adopted to solve the compactness
problem~\cite{090926A}.\\

\begin{figure}[!t]
  \centering
  \includegraphics[width=.45\linewidth]{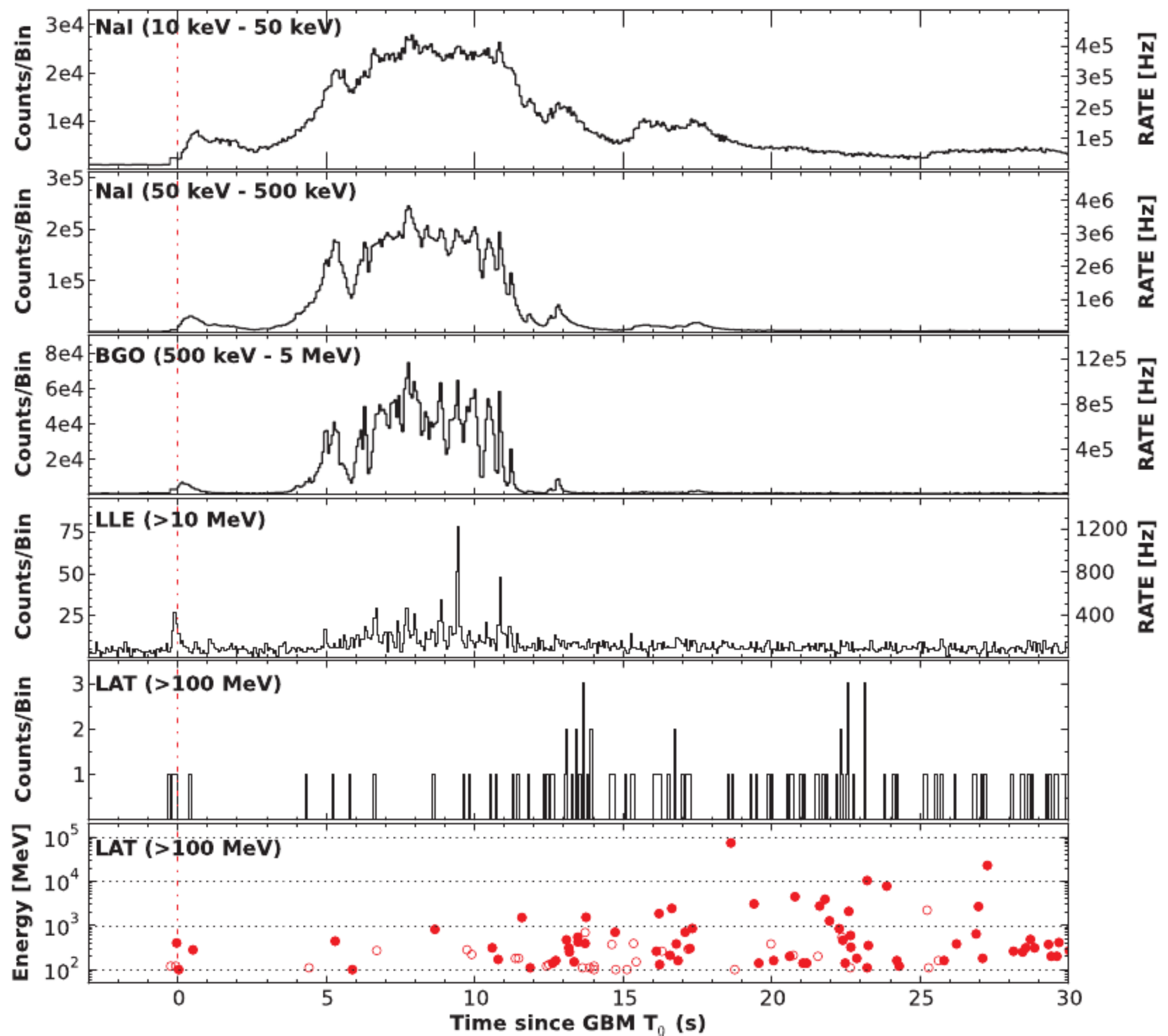}
  \includegraphics[width=.45\linewidth]{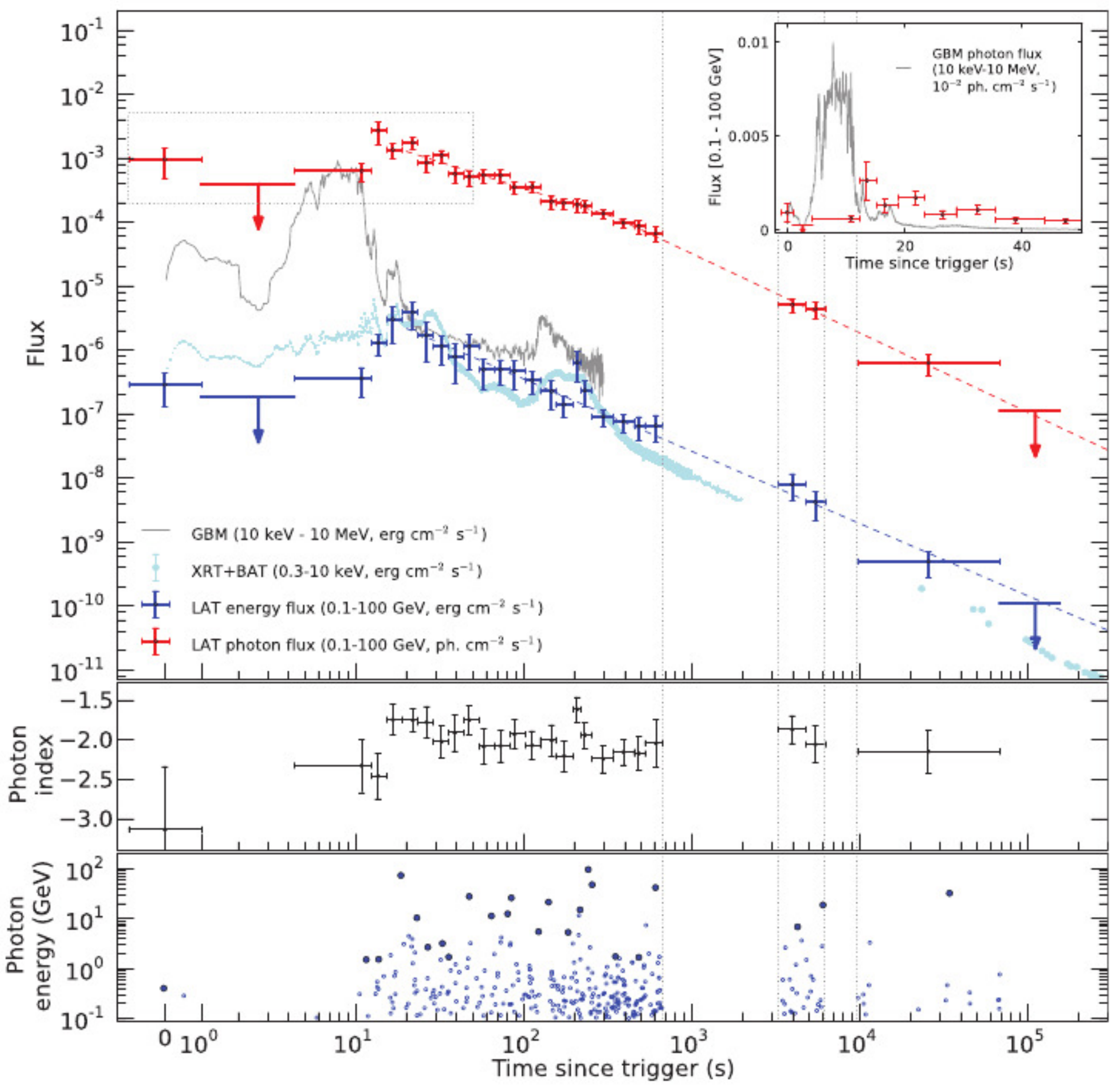}
  \caption{
    Left~: GRB\,130427A light curves as measured by the \Fermi/GBM and the \Fermi/LAT~\cite{130427a_lat} (see caption of
    figure~\ref{fig:090510}, left).
    Right~: GRB\,130427A energy flux light curves (in erg\,cm$^{-2}$\,s$^{-1}=10^{-3}$\,W\,m$^{-2}$) as measured
    by \Fermi (GBM and LAT instruments) and \Swift (BAT and XRT instruments)~\cite{130427a_lat}.
    The second panel shows the spectral index $\alex$ of the high-energy power law, and the last panel displays the energies of
    the individual photons which are well-reconstructed by the LAT above 100\,MeV.
  }
  \label{fig:130427a}
\end{figure}

GRB\,130427A is the brightest burst\footnote{The brightness of a burst is defined as the integral of its photon spectrum
  over time and energy or, equivalently, as the overall number of detected photons.} ever observed by
\Fermi~\cite{130427a_lat}, especially due to its proximity ($z=0.34$),
with more than 500 photons detected by the LAT above 100\,MeV, including more than 15 photons above 10\,GeV.
This burst has also the highest gamma-ray fluence\footnote{The fluence of a burst is defined as the integral of its energy
  flux over time and energy, thus it is indicative of its hardness:
  $\displaystyle F=T_{90}\int E\,\frac{dN}{dE}\,dE$, where $T_{90}$ is the burst duration, usually defined as the interval between the times where the burst has reached 5\% and 95\% of its total brightness in the same energy range.
} ever (larger than
$10^{-6}$\,J\,m$^{-2}=10^{-3}$\,erg\,cm$^{-2}$).
The high-energy emission started $\sim$10\,s after the keV--MeV emission, and it includes a 73\,GeV photon
detected 19\,s post-trigger.
Unlike other LAT bright GRBs, the high-energy emission of GRB\,130427A was very weak in the first instants and thus
appeared essentially uncorrelated with its keV--MeV emission (figure~\ref{fig:130427a}, left).
Most importantly, the extra power-law component of GRB\,130427A becomes significant only after the end of the GBM-detected emission.
As a result, most of the high-energy emission from this burst has been associated with the forward shock in the early afterglow
phase~\cite{130427a_lat}.
GRB\,130427A is also spectacular at later times (figure~\ref{fig:130427a}, right).
It has the brightest afterglow emission ever detected in X rays (though it is not the brightest intrinsically), as well as the
longest-lived emission at high energies, which lasted $\sim$19 hours.
Very energetic and late gamma rays have been recorded, in particular a 95\,GeV photon detected 244\,s post-trigger and a 32\,GeV
photon detected after 34.4\,ks. The former is the most energetic photon ever observed from a GRB, and the latter
dethrones the 18\,GeV photon detected by CGRO/EGRET after 4.5\,ks from GRB\,940217 (see section~\ref{sec:egret}).

\subsection{GRB common properties at high energies}
\label{sec:grbprop}
\begin{figure}[!t]
  \centering
  \includegraphics[width=.49\linewidth]{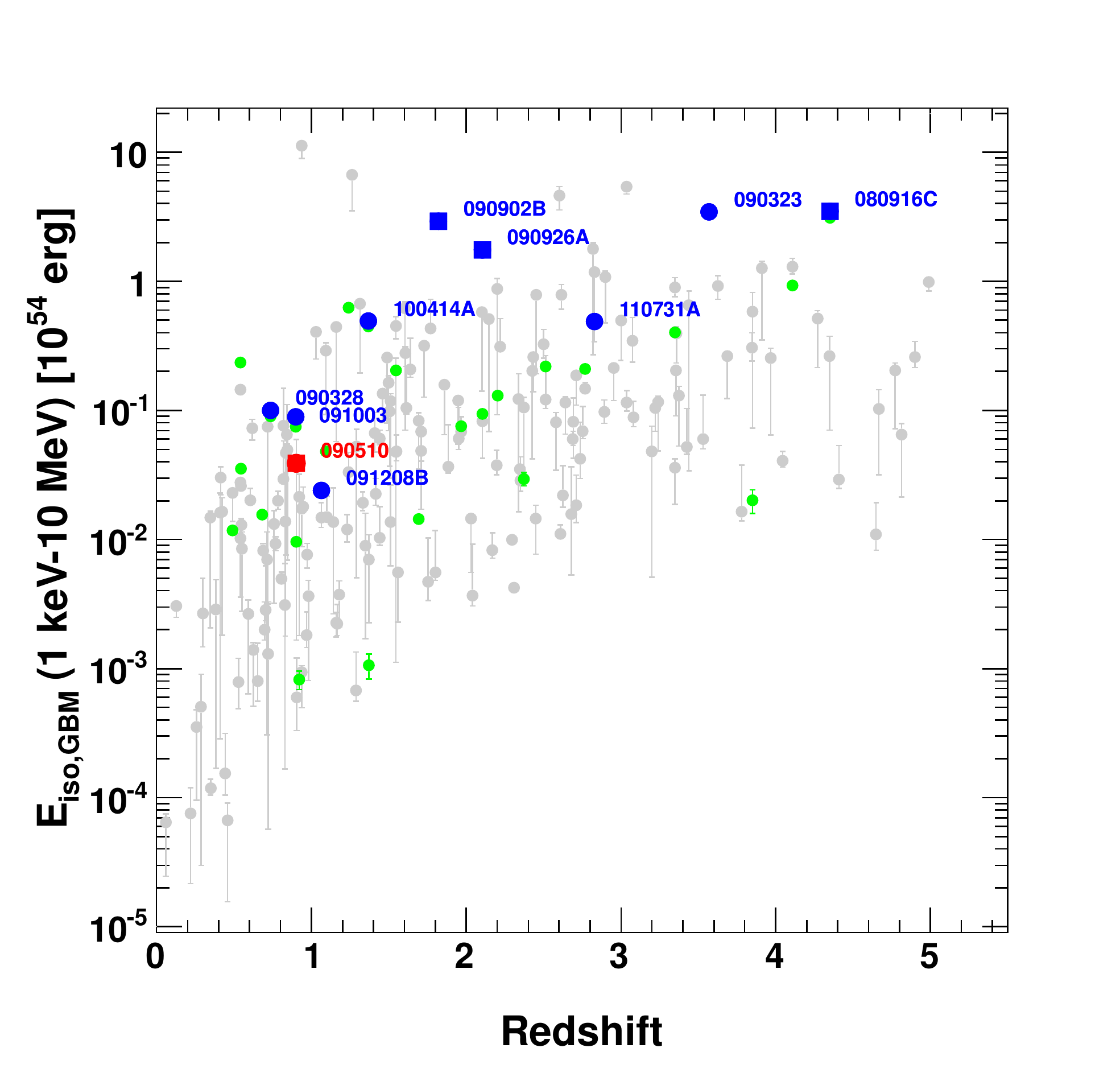}
  \includegraphics[width=.49\linewidth]{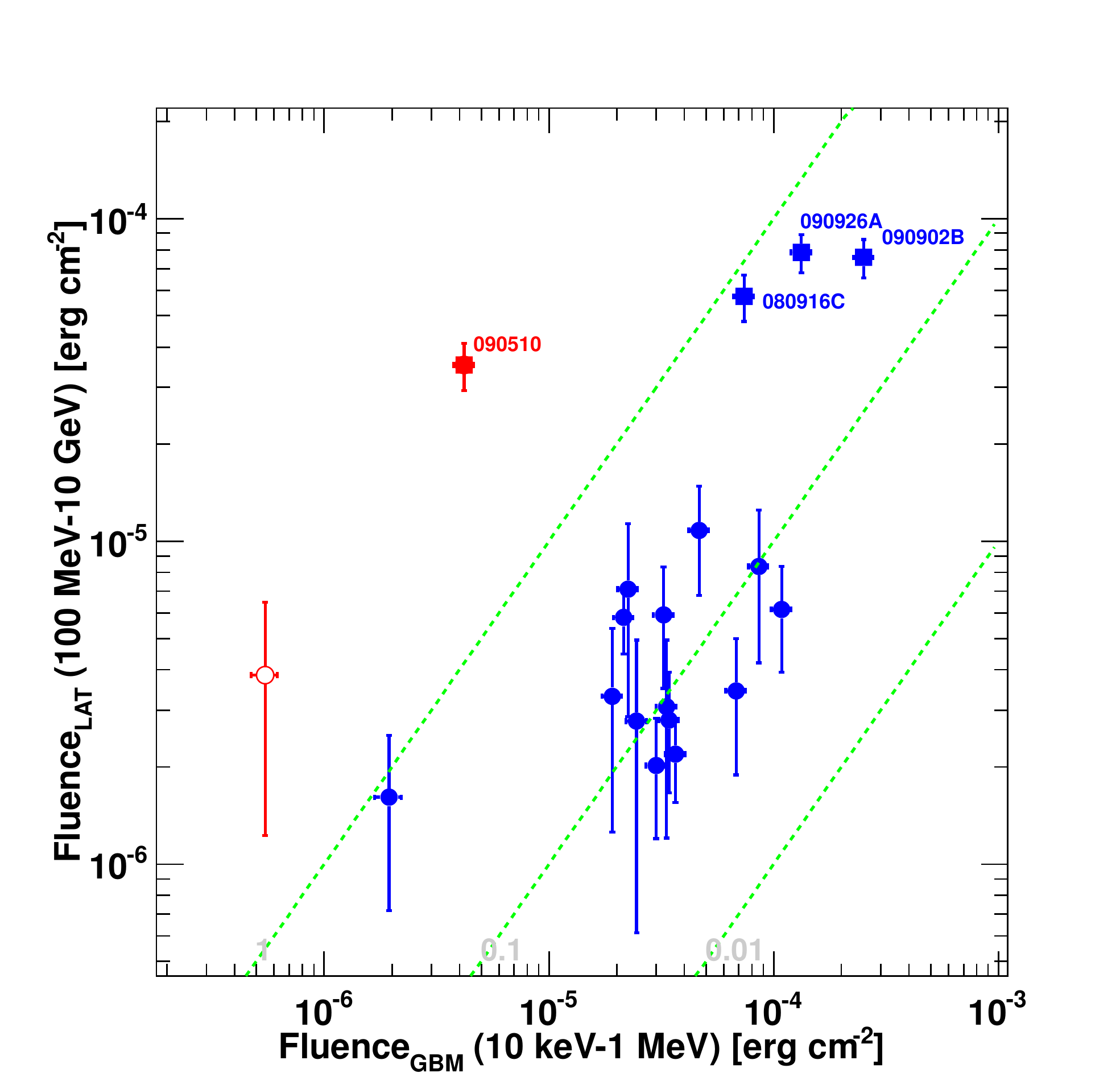}
  \caption{
    Left~: Isotropic equivalent energy $\eiso$  (in erg $=10^{-7}$\,J) in the source frame energy range $[E'_1, E'_2]=$~[1\,keV, 10\,MeV] as a function of the
    redshift for all the GRBs with measured distance in the first \Fermi/LAT catalog~\cite{grbcat},
    compared to the GRBs in the \Swift/BAT~\cite{butler2007} (in grey) and \Fermi/GBM~\cite{goldstein2012} (in green) catalogs.
    Right~: Fluence (in erg\,cm$^{-2}=10^{-3}$\,J\,m$^{-2}$) as measured by the \Fermi/GBM (10\,keV--1\,MeV) vs. that measured by the \Fermi/LAT ($>$100\,MeV) during their respective
    T$_{90}$'s (figure~\ref{fig:duration}, right) for all GRBs in the first \Fermi/LAT catalog~\cite{grbcat}.
    In both panels, long GRBs are displayed in blue, and short GRBs in red.
  }
  \label{fig:energetics}
\end{figure}

The first LAT GRB catalog~\cite{grbcat} is a systematic study of GRBs at high energies, covering the first three years of
\Fermi science operations.
Among the 733 GRBs detected by the GBM during this period, the LAT detected 35 bursts including 5 short bursts.
Not surprisingly, the LAT-detected GRBs were found to be among the GBM brightest and most fluent GRBs, due to the steepness
of the photon spectra and to the LAT limited effective area.
The complementary analysis reported in~\cite{racusin2011} showed that LAT-detected GRBs also have the brightest X-ray
luminosities in the afterglow phase (figure~\ref{fig:extended}, left).
In addition, they are among the most energetic GRBs (intrinsically and observationally), GRB\,090510 being the most
energetic short one (figure~\ref{fig:energetics}, left).
No particular trend in redshift with respect to the \Fermi/GBM and \Swift/BAT samples was found.
Whereas the fluence in the LAT energy range amounts only to $\approx10$\% of the fluence in the GBM energy range for long GRBs,
short GRBs seem to have a much larger fluence ratio (figure~\ref{fig:energetics}, right).
Although this result certainly requires more GRB statistics to be firmly confirmed, it suggests different energy outputs above
100\,MeV from GRBs depending on their progenitors.
Apart from this property, short and long GRBs have shown very similar characteristics in the behaviour of their high-energy
emission.\\
\begin{figure}[!t]
  \centering
  \includegraphics[width=.49\linewidth]{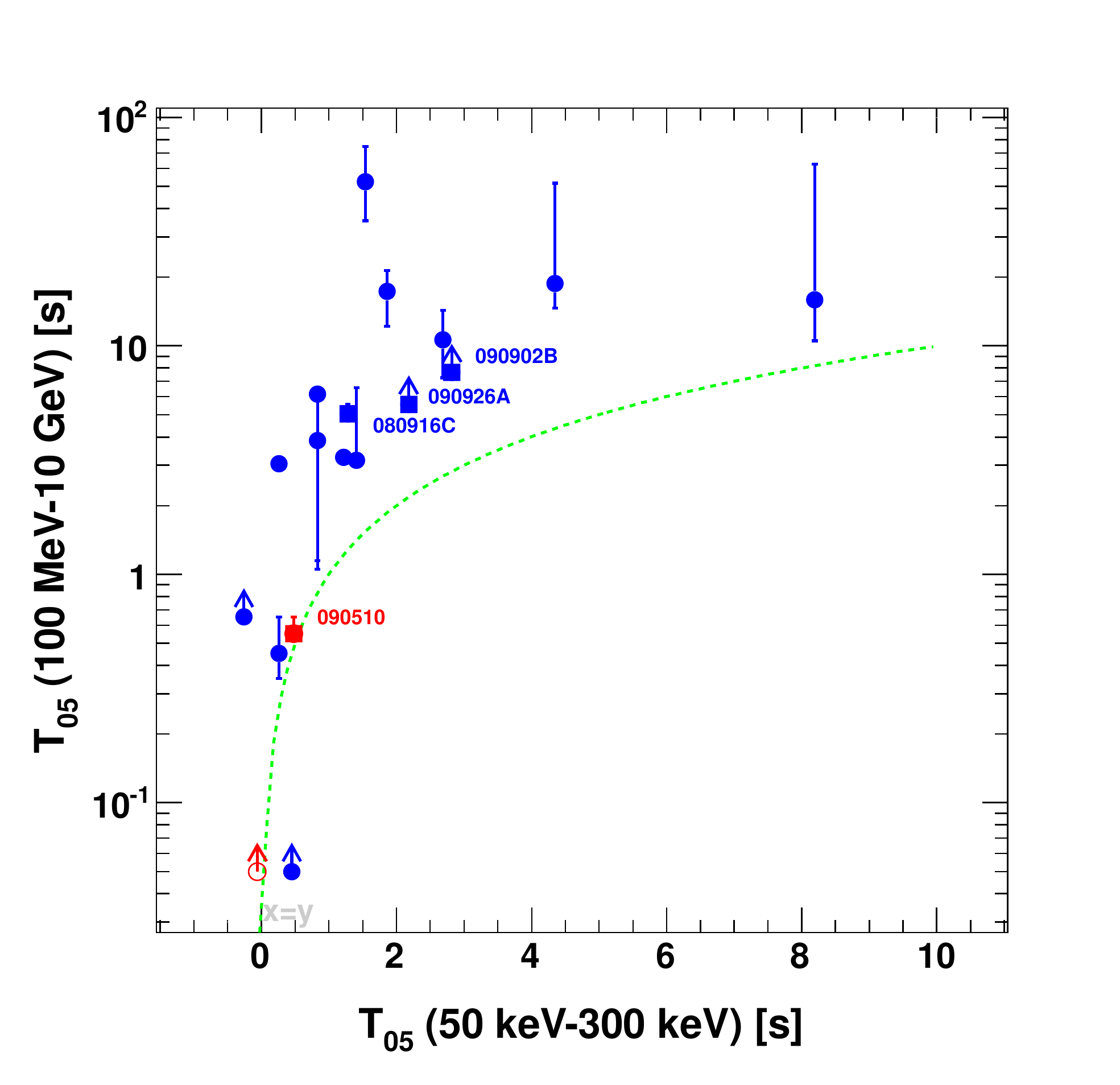}
  \includegraphics[width=.49\linewidth]{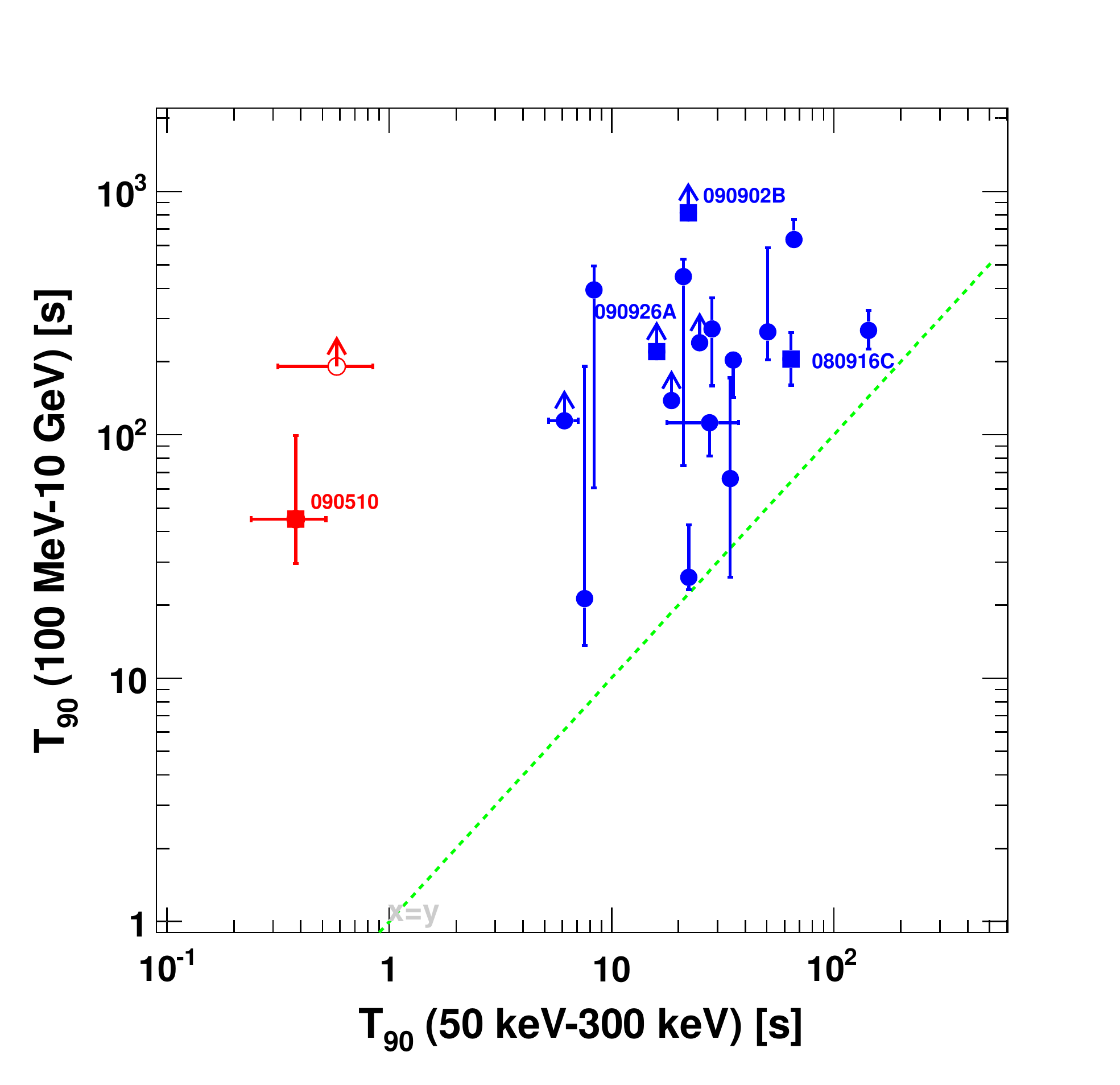}
  \caption{Onset time (T$_{05}$, left panel) and duration (T$_{90}$, right panel) of the GRB emission as measured by the \Fermi/GBM
    (50--300\,keV) vs. that measured by the \Fermi/LAT ($>$100\,MeV) in the first \Fermi/LAT catalog~\cite{grbcat}.
    In both panels, long GRBs are displayed in blue, and short GRBs in red.
  }
  \label{fig:duration}
\end{figure}

The GRB emission detected by the LAT above 100\,MeV is systematically delayed
with respect to the emission observed with the GBM at keV--MeV energies (figure~\ref{fig:duration}, left), and it lasts
systematically longer (figure~\ref{fig:duration}, right).
Specifically, the first pulses in the keV--MeV light curves have no simultaneous counterparts, or only much weaker ones, at high energies, while later pulses can coincide in different energy bands (figures~\ref{fig:090510} and~\ref{fig:090926a}, left).
In other words, this delay is not caused by an overall shift of the high-energy temporal structures with respect to their
potential keV--MeV counterparts.\\

Joint spectral fits using \Fermi GBM and LAT data recorded during the prompt phase showed that the commonly used Band function
does not capture all spectral characteristics.
Among the 30 GRBs analysed in~\cite{grbcat}, this phenomenological shape alone can reproduce the spectrum of 21 GRBs, all other
spectra but two (GRBs\,090626 and 110721A, for which a logarithmic parabola alone is preferred) are best fitted using either an
extra power-law component (GRBs\,080916C, 090510, 090902B, 090926A, 100414A and 110731A) and/or an exponential cutoff
(GRBs\,090926A and 100724B) at high energies.
Other deviations from the Band function have been found at lower energies as well. Some GRB spectra contain an important flux in
excess to the Band function below $\sim$50\,keV, as reported for GRBs\,090510 and 090902B in section~\ref{sec:grbind}.
Moreover, several GRB spectra (e.g., GRBs 100724B, 110721A and 120323A) exhibit a shoulder on top of the low-energy branch of the
Band function, which has been interpreted~\cite{100724b,110721a,120323a} as the jet thermal emission expected at the photospheric
radius ($R_\mathrm{ph}\approx10^{9-10}$\,m, where the jet plasma becomes transparent to its own radiation)~\cite{daigne2003}.
These examples illustrate the GRB spectral diversity in their prompt emission phase, and clearly call for the development of detailed
broad-band physical models as a prerequisite to understanding GRB emission at the highest energies (see section~\ref{sec:grbint}).\\

At late times, the GRB high-energy emission looks simpler and its origin seems less unclear.
After the end of the keV--MeV prompt emission, no noticeable spectral evolution is observed, the photon spectrum is well
reproduced by a single power-law component of index $\alex\simeq-2$ (see, e.g., the second panel of figure~\ref{fig:130427a},
right), and the luminosity decays smoothly (figure~\ref{fig:extended}, right).
As discussed in section~\ref{sec:grbint}, such a decay phase is consistent with a forward shock origin of the high-energy
emission during the afterglow phase.
Specifically, the high-energy luminosity decays as $\displaystyle L(t)\propto t^{\alt}$ with $\alt\simeq-1$ for all GRBs but three (GRBs\,090510,
090902B and 090926A), for which a broken power law is detected~\cite{grbcat}.
For these bursts, the time of the break is observed after the end of the keV--MeV emission as measured by the GBM, and
$\alt$ switches from $\approx-2$ to $-1$.
Defining the decay index $\all$ as the value of $\alt$ at very late times, all GRBs\footnote{Except GRBs\,080916C and
  110731A. However, these long GRBs have the shortest intrinsic durations at high energies, suggesting that a non-detected temporal
  break could still be present at later times.} thus follow a unique power-law decay with $\all\simeq-1$.

\section{Physical implications and discussion}
\label{sec:grbphys}
\subsection{Constraints on GRB jet Lorentz factors}
\label{sec:grbgamma}
\begin{figure}[!t]
  \centering
  \includegraphics[width=.54\linewidth]{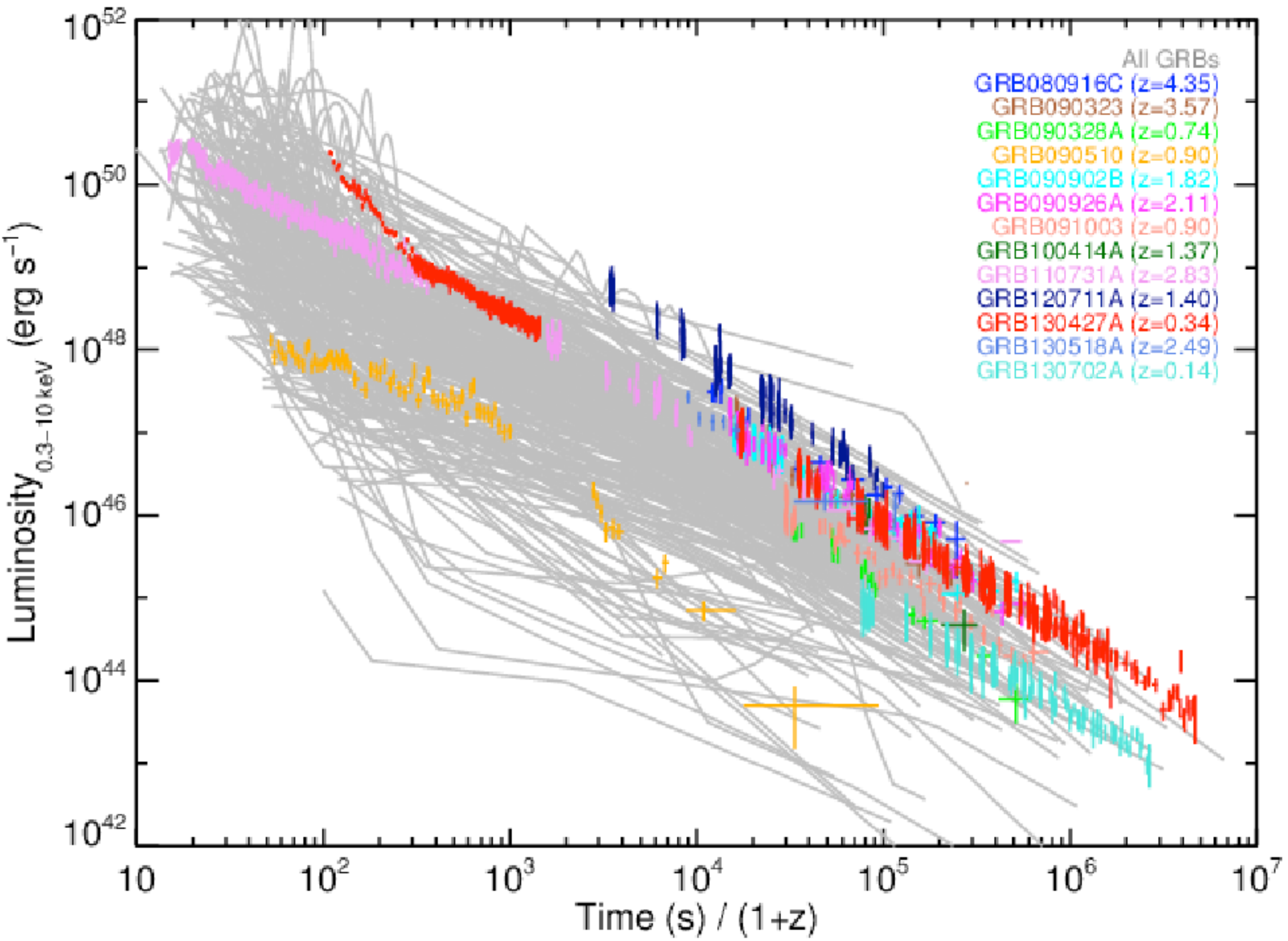}
  \includegraphics[width=.45\linewidth]{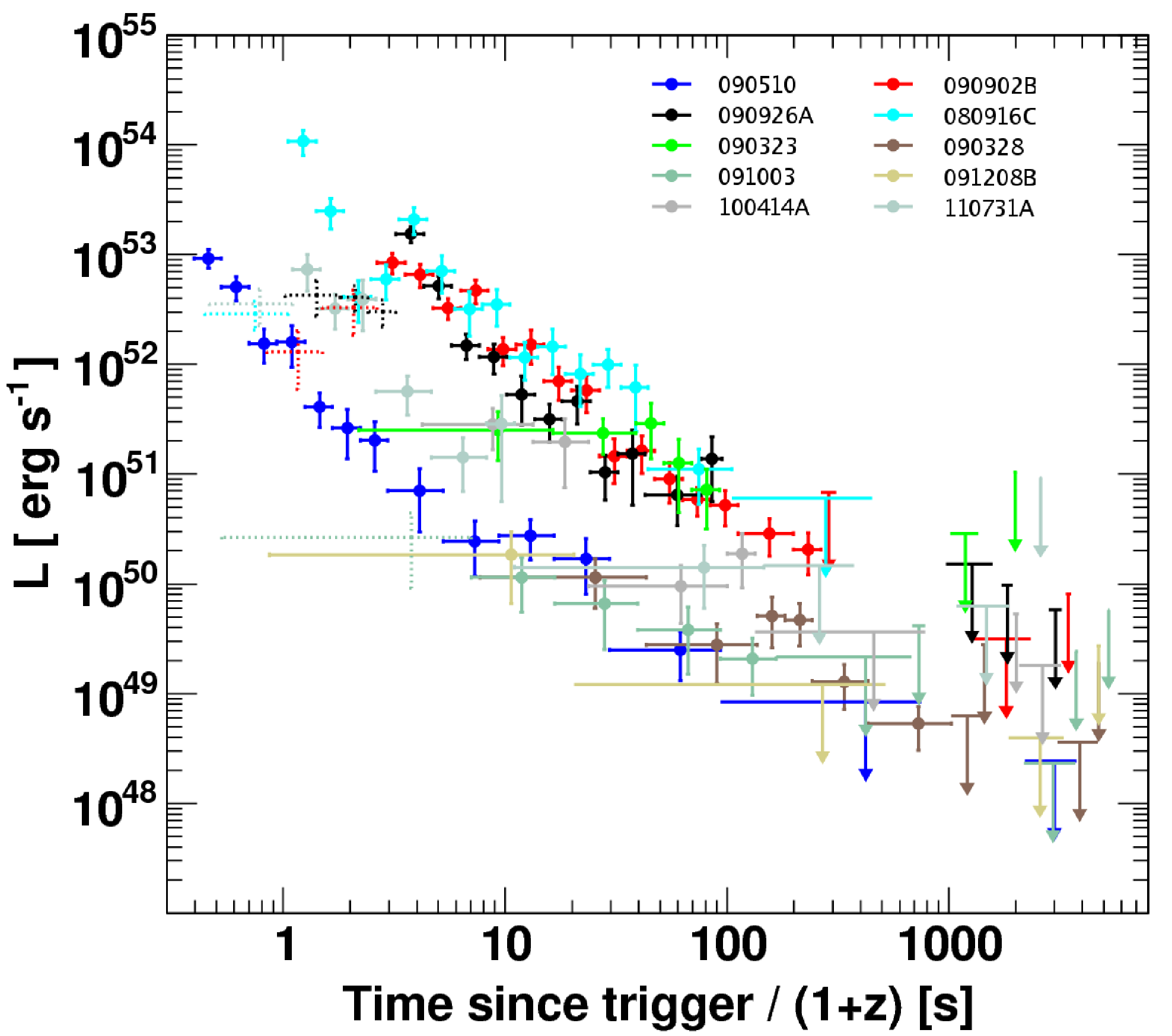}
  \caption{
    Left~: Decay of the X-ray (0.3--10\,keV) luminosity (in erg\,s$^{-1}=10^{-7}$\,W) with time measured in the
    source frame for two samples of GRBs detected
    by \Swift (in grey) and \Fermi/LAT (in color)~\cite{racusin2011}.
    Right~: Decay of the high-energy ($>$100\,MeV) luminosity (in erg\,s$^{-1}=10^{-7}$\,W) with time measured
    in the source frame for all the GRBs with known
    redshift and detected extended emission in the first \Fermi/LAT catalog~\cite{grbcat}.
  }
  \label{fig:extended}
\end{figure}

The \Fermi/LAT has detected high-energy emission without any spectral attenuation beyond a few (tens of) GeV from several bright
GRBs~\cite{080916C,090510_prompt,090902B,090926A}.
Setting the gamma-ray opacity to pair production to unity in equation~\ref{eq:opacity}, the highest-energy detected photons
provided lower limits
$\Gamma_\mathrm{min}\approx1000$ on the jet Lorentz factor (figure~\ref{fig:gamma}, left), revealing that both long and short GRBs
can have high outflow velocity, a key result for GRB modeling.
It must be noted that the identification of the soft photons responsible for the $\gamma\gamma$ absorption is a delicate
task, and ideally requires spectroscopy with good statistics over the considered variability time scale $t_v$ (see
equation~\ref{eq:opacity}).
In pratice, $t_v$ is chosen as the fastest variability observed in the \Fermi/GBM light curve (e.g., 50\,ms for GRB\,090902B, while some
variability is observed in the LAT down to $\sim$90\,ms~\cite{090902B}), and the target photon spectrum is derived over slightly
larger durations.
Moreover, the target photon field is considered uniform, isotropic and time-independent in this simple one-zone steady-state model.
More realistic computations~\cite{granot08,hascoet2012}, which account for geometrical and dynamical effects, can lead to smaller
opacities and thus to $\Gamma_\mathrm{min}$ values which are 2 to 3 times smaller.\\

Whereas $\sim$9.3 LAT-detected GRBs with more than 10 photons above 100\,MeV were expected per year from
pre-launch estimates~\cite{band09}, a mean rate of $\sim$6.3 GRBs per year was obtained in the first \Fermi/LAT
catalog~\cite{grbcat}.
The past estimates were possibly affected by systematic uncertainties, e.g., arising from the extrapolations from the CGRO/BATSE
energy range to the LAT energy range, which are uncertain due to the large lever arm combined with the errors $\epsilon_\beta$ in
measuring the Band function index $\beta$.
However, the lack of LAT GRB detections raises the question whether the high-energy emission is suppressed and if spectral cutoffs
are more common than anticipated, similarly to the attenuated spectrum of GRB\,090926A (figure~\ref{fig:090926a}, right).
Such cutoffs should be even more pronounced and affect the observed spectra above a few tens of MeV in order
to completely turn off the high-energy emission and to prevent any high-energy component from emerging in the LAT energy range.\\

In order to investigate this question, a list of LAT non-detected GRBs which were bright and/or spectrally hard in the GBM energy domain
has been extracted from a sample of 288 GRBs detected by the GBM during the first 2.5 years of \Fermi science operations~\cite{latul}.
Among the 30 GRBs with more than 70 counts/s in the GBM BGO detectors and with a good measurement of the Band function index
$\beta$ ($\epsilon_\beta<0.5$), 6 GRBs require a degree of spectral softening between the BGO and the LAT energy ranges to explain
their LAT non-detection.
The flux and hardness ($E_P$ and $\beta$, see equation~\ref{eq:band}) of these GRBs are representative of the initial sample.
However, they have the smallest $\epsilon_\beta$ values, indicating that a very accurate spectroscopy is required to reveal the
spectral feature.
Assuming that this softening is due to gamma-ray opacity to pair production, upper limits $\Gamma_\mathrm{max}$ ranging from 150
to 650 were derived for the jet Lorentz factors of these GRBs, using a similar formalism as in equation~\ref{eq:opacity} with a
conservative $t_v=$100\,ms variability time scale.
As a result, the comparison of these limits to previous constraints on $\Gamma$ suggests that GRBs span a relatively broad range
of jet velocities (figure~\ref{fig:gamma}, left).

\subsection{Possible origins of the GRB emission at high energies}
\label{sec:grbint}
\begin{figure}[!t]
  \centering
  \includegraphics[width=0.48\linewidth]{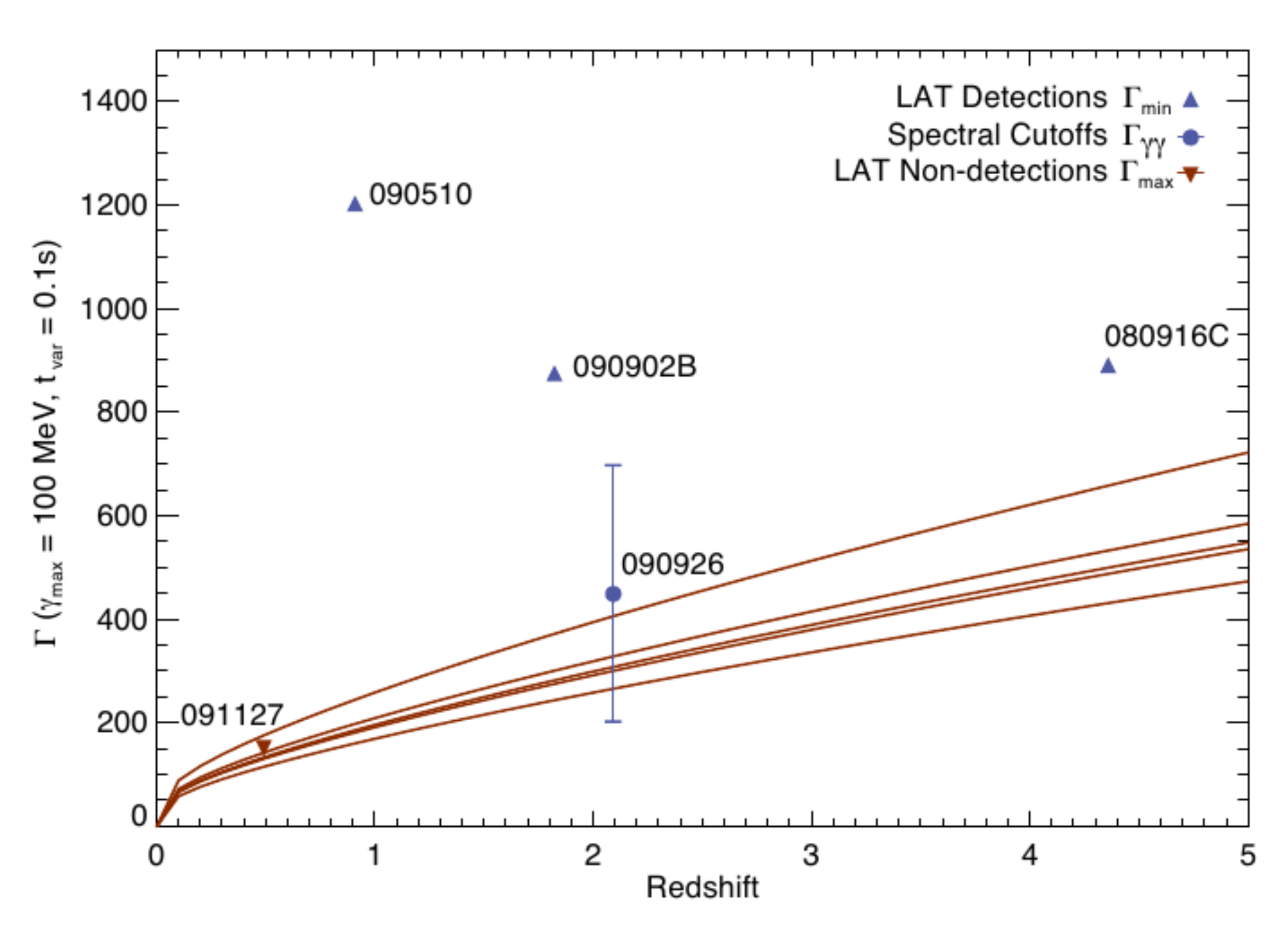}
  \includegraphics[width=0.50\linewidth]{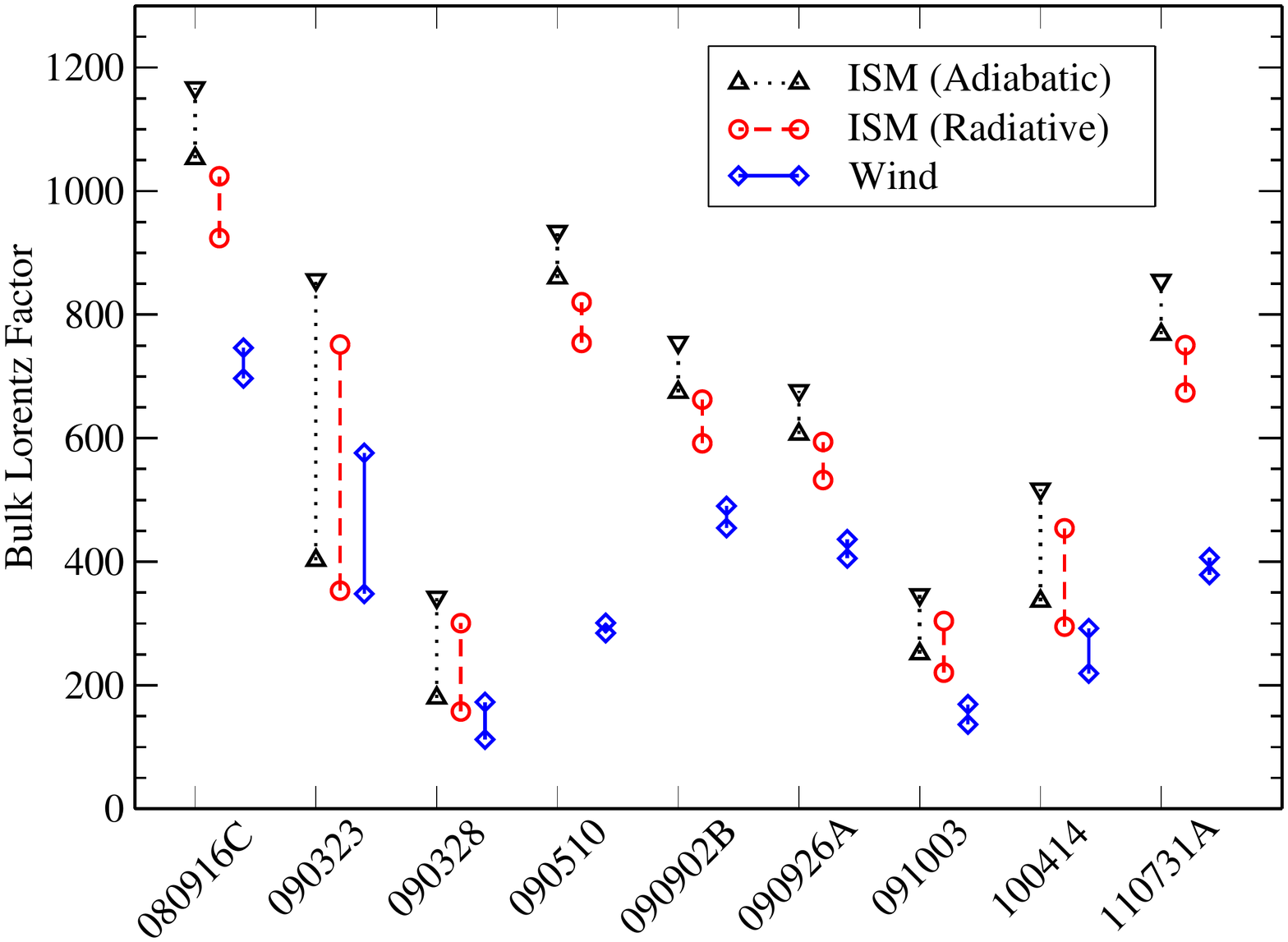}
  \caption{
    Constraints from \Fermi on GRB jet Lorentz factors.
    Left~: Curves giving upper limits $\Gamma_\mathrm{max}$ on the jet Lorentz factors as a function of the redshift for 6 GRBs
    exhibiting a spectral softening at a few tens of MeV~\cite{latul}: GRBs\,080925, 081207, 090131, 090528B, 100724B, and
    091127. The latter burst has a known redshift and is shown with a brown triangle.
    Lower limits $\Gamma_\mathrm{min}$ (GRBs\,080916C, 090510, 090902B) and measurement (GRB\,090926A) for \Fermi/LAT bright GRBs are
    superimposed (blue triangles and point, respectively).
    The target photon field for $\gamma\gamma$ absorption is assumed uniform, isotropic and time-independent, but the error bar
    for GRB\,090926A accounts for different models~\cite{090926A}, illustrating the overall scaling that could be applied to the
    entire figure.
    Right~: Jet Lorentz factor for all the GRBs with known redshift in the first \Fermi/LAT catalog~\cite{grbcat}.
    These estimates assume that the high-energy peak flux time marks the start of the jet deceleration by the circum-burst medium
    (inter-stellar medium with uniform density or massive star wind, which is more appropriate for long GRBs).
  }
  \label{fig:gamma}
\end{figure}

{\it Emissions of internal origin}\\

From an observational point of view, the GRB prompt phase can be reasonably defined as the period of time where the
keV--MeV emission consists of impulsive and pulsed structures. 
This definition might be too simplistic since it relies on durations (i.e., $T_{90}$'s) obtained with background-limited
detectors such as the \Fermi/GBM, but I will adopt it in the following discusion.
The GRB prompt emission in the keV--MeV domain is often attributed to the synchrotron emission of internal-shock accelerated electrons.
At low energies, the synchrotron model predicts a spectral index value between $-$3/2 (for electrons in a fast-cooling regime) and
$-$2/3 (in a slow cooling regime)~\cite{sari1998}, whereas the observed Band function index $\alpha$ is distributed around $-1$.
The GRB extreme variability and brightness imply efficient radiative processes and suggest that the slow-cooling regime is not
suitable. Anyhow, many GRBs have shown to be hard enough at low energies for their Band function index $\alpha$ to exceed the limit in this regime as well~\cite{preece1998}.
Since this so-called line-of-death problem challenges the internal shock synchrotron model, several theoretical extensions have been recently proposed which predict, e.g., an increase of the spectral index caused by the
retreatment of low-energy photons through inverse Compton processes~\cite{daigne2011} or by the presence of an additional thermal
component~\cite{100724b,110721a,120323a}, or a pure thermal emission in extreme cases like GRB\,090902B~\cite{ryde10}.\\

At high energies, two different classes of models consider the emission of internal-shock accelerated particles in order to explain
the extra power-law component observed in the prompt phase of \Fermi/LAT bright GRBs (a~priori, these models do not address the
temporal extension of the high-energy emission).
So-called leptonic models are based on electron synchroton emission at keV--MeV energies and inverse Compton or SSC processes at high
energies.
They naturally predict the fast variability observed in GRB light curves, and the temporal correlation between different energy
bands, similarly to the sharp pulse of GRB\,090926A.
However, this first class of models needs some fine tuning (see, e.g.,~\cite{bosnjak2009}) to produce a delayed onset of the
high-energy emission which is longer than the spike widths, each pulse in the light curve marking a different shell collision and shock.
Such models also have difficulties to produce the flux excess which is sometimes observed below $\sim$50\,keV (figure~\ref{fig:090510},
right).
This flux excess is indeed compatible with the extension to the lowest energies of the high-energy extra power-law component,
which is not expected from inverse Compton processes.\\
\begin{figure}[!t]
  \centering
  \includegraphics[width=.49\linewidth]{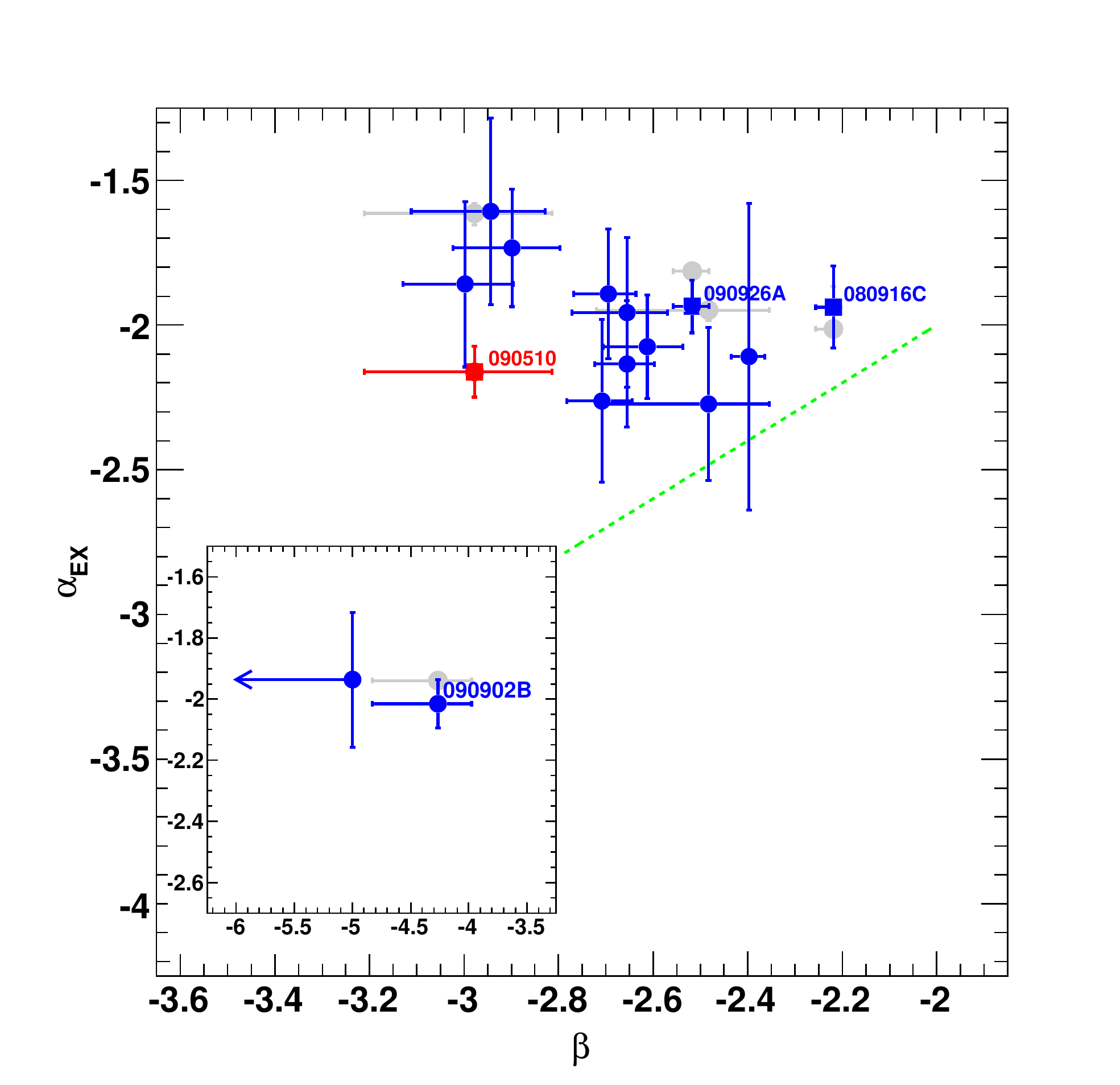}
  \parbox[h]{0.38\linewidth}{
    \vspace{-7cm}
    \caption[]{
      Comparison, for all the GRBs analysed in the first \Fermi/LAT catalog~\cite{grbcat}, of the Band function index $\beta$ with the
      spectral index $\alex$ of the high-energy power-law component.
      Values displayed in color (blue for long GRBs and red for the short GRB\,090510) were obtained from a power-law spectral fit
      (yielding $\alex$) to the LAT data recorded after the end of the GBM-detected emission, and from a Band spectral fit
      (yielding $\beta$) to the GBM and LAT data recorded during the period of the GBM-detected emission (T$_{90}$).
      If, in the latter case, the data require an extra power law beyond the Band function at high energies, then both fitted
      indices are indicated in grey.
    }
\label{fig:betagamma}
}
\end{figure}

So-called hadronic models, which form the second class of models, investigate GRBs as possible sources of the ultra-high energy cosmic
rays~\cite{vietri1997} (of energies greater than 1\,EeV=$10^{18}$\,eV)\footnote{The secondary neutrino emission implied by
  hadronic models is potentially detectable by future experiments such as IceCube ({\tt{http://icecube.wisc.edu}}) and KM3NeT ({\tt{http://www.km3net.org/home.php}}).
GRBs are thus good source candidates for the development of multi-messenger astrophysics in the coming years and for the
advance of high-energy neutrino astronomy, which is still in its infancy~\cite{icecube}.}.
In the high-energy gamma-ray domain, hadronic models consider the proton synchrotron emission~\cite{razzaque10a,razzaque10b}
and/or the inverse Compton emission from secondary $e^+e^-$ pairs produced in internal cascades initiated by accelerated protons
or ions (through $p+\gamma\rightarrow p/n+\pi^0/\pi^+$ processes)~\cite{asano09}.
In these models, the delayed onset of the high-energy emission could result from the time needed to accelerate protons and ions
and to develop cascades, but these models do not naturally predict the aforementionned correlated variability and have
difficulties to produce high-energy pulses with similar width as the keV--MeV pulses.
While synchrotron emission from secondary $e^+e^-$ pairs could explain the flux excess below
$\sim$50\,keV, these models require an energy injected in the magnetic fields or in the protons which is $10^{2-3}$ larger than
observed.
It must be noted that in the proton synchrotron model, this constraint strongly depends on the jet Lorentz
factor ($\propto\Gamma^{16/3}$, see~\cite{razzaque10a}) and could be accommodated with low values of this key parameter similar
to those discussed in section~\ref{sec:grbgamma}.\\

{\it Emissions of external origin}\\

An alternative interpretation of GRB properties at high energies, which is nevertheless also not fully satisfactory, is provided by the
early and late afterglow models.
Unlike the internal shock models discussed above, these models consider the synchroton emission from electons accelerated at the
forward shock, which is produced by the interaction of the jet with the circum-burst medium.\\

The smooth decay of the high-energy luminosity at late times (see section~\ref{sec:grbprop} and figure~\ref{fig:extended}, right) is
similar to the behavior of the visible/UV and X-ray emission in the afterglow phase (figure~\ref{fig:extended}, left).
In addition, figure~\ref{fig:betagamma} shows that the spectral index $\alex$ of the extra power-law component is clustered around
$-2$, either after the end of the keV--MeV emission (points in color in the figure; see also the second panel of
figure~\ref{fig:130427a}, right) or during the prompt phase of \Fermi/LAT bright GRBs (grey points; see also figures~\ref{fig:090510} and~\ref{fig:090926a}, right).
This stability of the high-energy spectrum contrasts with the spectral variability of the keV--MeV emission, as shown in figure~\ref{fig:betagamma} by the different values of the Band function index $\beta$.
Finally, $\beta$ is systematically lower than $\alex$.
These findings suggest that the high-energy emission results essentially from a mechanism which is independent from internal
shocks, namely from the forward shock.
Furthermore, they indicate that the forward shock high-energy emission not only accounts for the luminosity observed during the
late afterglow phase, and that it can also be dominant while the prompt keV--MeV emission from internal shocks remains detectable.
It must be stressed, however, that the high-energy contribution from internal shocks is still required during
highly variable episodes in the prompt phase, as in the case of the sharp pulse of GRB\,090926A.
Distangling both contributions may be possible in the future through time-resolved spectral analyses of \Fermi data which
combine spectroscopy with a full characterization of the variability properties at high energies.
Future LAT-detected bursts similar to GRBs\,090510, 090902B and 090926A would be also helpful, since the initial steep decay
($\alt\approx-2$) observed in their high-energy light curves may mark the transition from a phase where the forward shock emission is
contaminated or even dominated by the prompt high-energy emission, to a pure afterglow phase.\\

The forward shock synchrotron emission models have been successfully applied to the \Swift and \Fermi
multi-wavelength observations of the afterglow phase of GRBs\,090510 and 110731A~\cite{090510_afterglow,110731A,lemoine2013,wang2013}.
These models predict a relation between the luminosity decay index $\alt$ and the spectral index $\alex$, which
depends on the considered energy range. Above 100\,MeV, electrons are in a fast-cooling regime and radiate efficiently.
In this case, the relation takes the form
$\alt=(14+12\,\alex)/7\simeq -10/7$ for a blast wave expansion where radiative losses are important, and
$\alt=(4+3\,\alex)/2\simeq-1$ for an adiabatic expansion~\cite{sari1998,granot2002}.
The latter case is clearly favored by the decay index $\all\simeq-1$ measured by the \Fermi/LAT at very late times (see
also~\cite{razzaque10b,kumar2009,nava2014}).\\
\begin{figure}[!t]
  \centering
  \includegraphics[width=.44\linewidth]{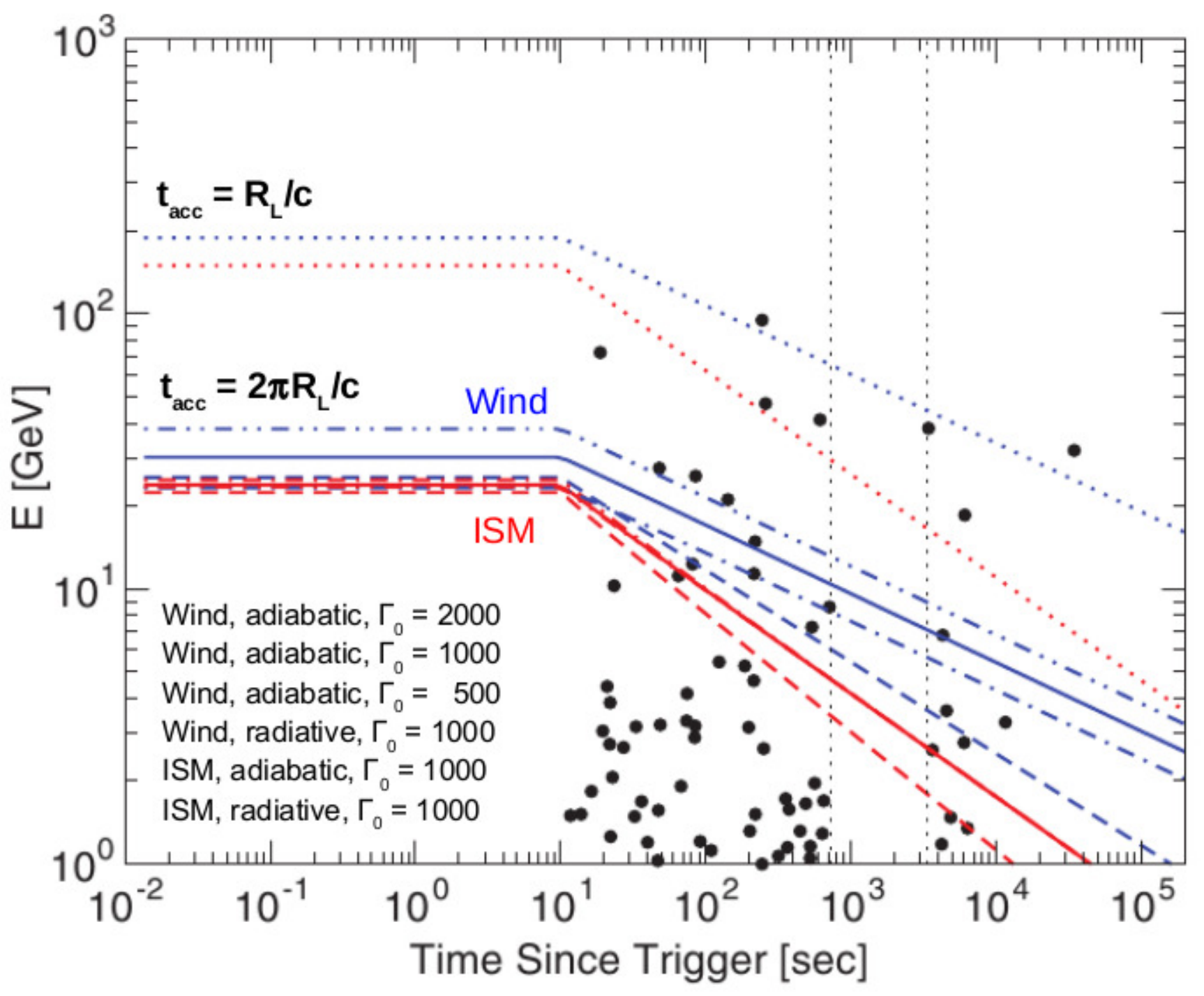}
  \includegraphics[width=.5\linewidth]{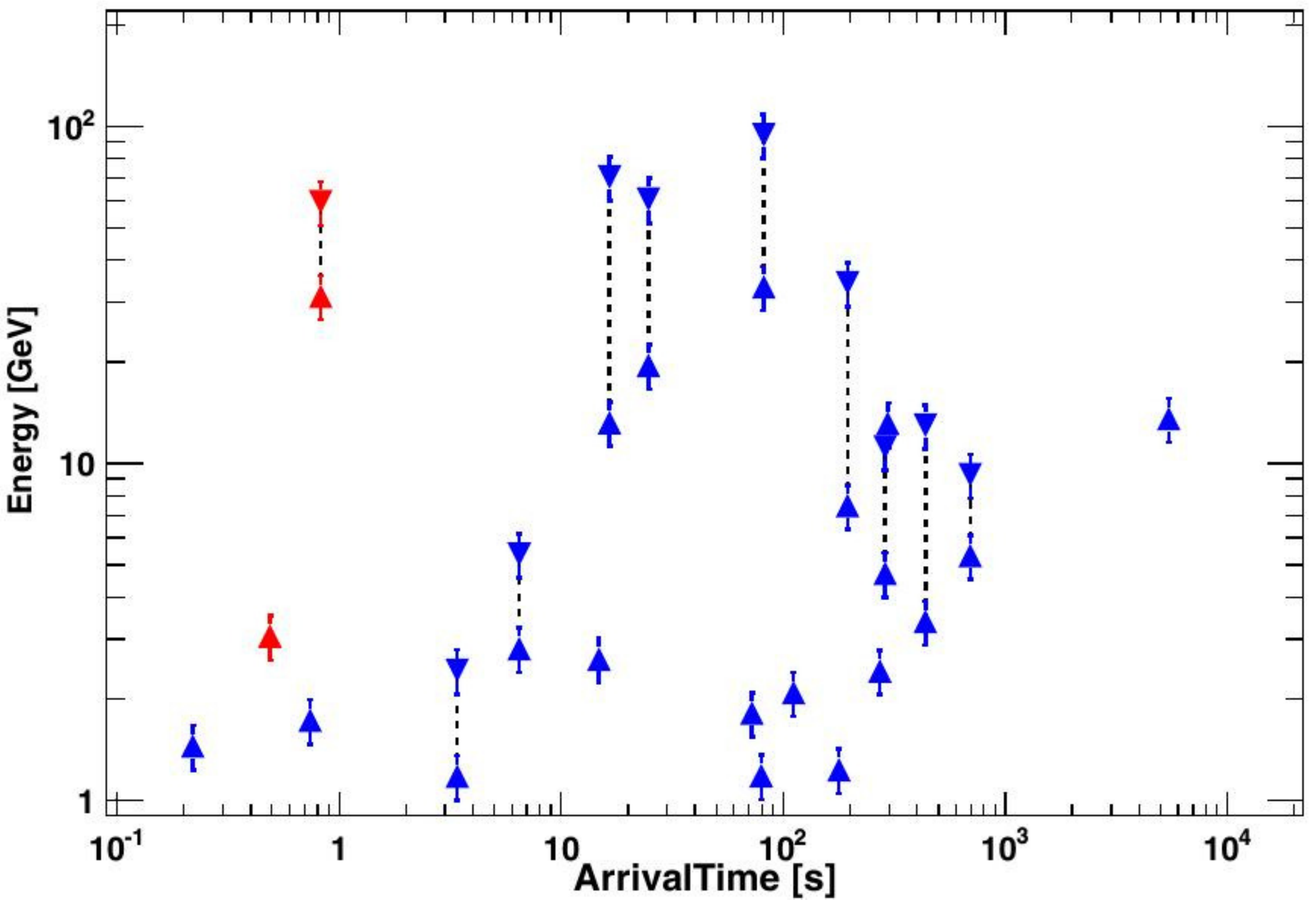}
  \caption{
    Left~: Maximum synchrotron energy in the foward shock model for GRB\,130427A as a function of time~\cite{130427a_lat}.
    Line styles denote different hypotheses regarding the initial value of the jet Lorentz factor and the structure of the
    circum-burst medium (inter-stellar medium with uniform density or massive star wind, which is more appropriate for long GRBs).
    Black dots indicate the high-energy photons detected by the \Fermi/LAT.
    Right~: Observed (upward-pointing triangles) and source frame (downward-pointing triangles) energy and arrival time for the
    highest-energy photons associated with the long (blue) and short (red) GRBs in the first \Fermi/LAT catalog~\cite{grbcat}
  }
  \label{fig:emax}
\end{figure}

The forward shock synchrotron emission models can also explain the delayed onset of the high-energy emission, which is attributed
to the time required for the flux to increase during the early afterglow phase until it becomes detectable by the
\Fermi/LAT~\cite{090510_afterglow,razzaque10b,kumar2009,ghisellini2010}.
The high-energy flux increases as $t^2$ due to the progressive energy dissipation which results from the jet deceleration by the
circum-burst medium.
After the peak flux time, the outflow slows down considerably and reaches non-relativistic velocities~\cite{blandford1976,sari1997}.
Estimates of the jet Lorentz factor $\Gamma$ can be obtained from the observation of the early afterglow phase, assuming that the peak
flux time which is measured by the LAT above 100\,MeV is of the order of the jet deceleration time.
These estimates are thus larger for smaller delays, and they range from 200 to 1000 (figure~\ref{fig:gamma}, right) for all the
GRBs with known redshift in the first \Fermi/LAT catalog~\cite{grbcat}.
These values are compatible with those obtained from opacity arguments (see section~\ref{sec:grbgamma}), and they provide an independent
confirmation that both long and short GRBs have relativistic outflows.\\

Afterglow models can be further explored at high energies by comparing the most energetic photons detected by the \Fermi/LAT with
the maximum synchrotron photon energy, $E_\mathrm{syn,max}$.
Assuming a single acceleration and emission region, this energy can be derived by equating the electron acceleration time scale,
$t_\mathrm{acc}$, and the electron energy loss time scale due to synchrotron radiation.
The acceleration time scale can be approximated as the inverse of the Larmor angular frequency ($t_\mathrm{acc}\simeq R_\mathrm{L}/c$,
which corresponds to an extremely fast acceleration) or as the Larmor time scale for an
electron to execute a gyration ($t_\mathrm{acc}\simeq 2\pi\,R_\mathrm{L}/c$, more realistic)~\cite{cras2015-theory}.
In the case of GRB\,130427A, this yields a conservative limit
$\displaystyle E_\mathrm{syn,max}(t)\approx80\,\Gamma(t)$\,MeV~\cite{130427a_lat}.
As shown in the left panel of figure~\ref{fig:emax}, this limit is violated by several LAT-detected photons.
In particular, the 95\,GeV photon detected after 244\,s and the 32\,GeV photon detected after 34.4\,ks are clearly incompatible
with a synchrotron origin.
Although an SSC interpretation has been considered by several authors~\cite{liu2013,fan2013}, this scenario is not favored by the absence of an inverse Compton spectral component in the LAT energy range~\cite{130427a_lat}, which is reinforced by joint detections
with the \Nustar X-ray satellite after 1.5 and 5 days~\cite{130427a_nustar}, and by flux upper limits
from  the VERITAS Cherenkov telescopes in the very high-energy range after 1 day~\cite{130427a_veritas}.
\Fermi/LAT observations of GRB\,130427A therefore challenge the forward shock synchrotron emission models in their simplest
version, and they call at least for a better description of external shock micro-physics.

\section{Prospects for GRB observation at very high energies}
\label{sec:vhe}
\begin{figure}[!t]
  \centering
  \includegraphics[width=.51\linewidth]{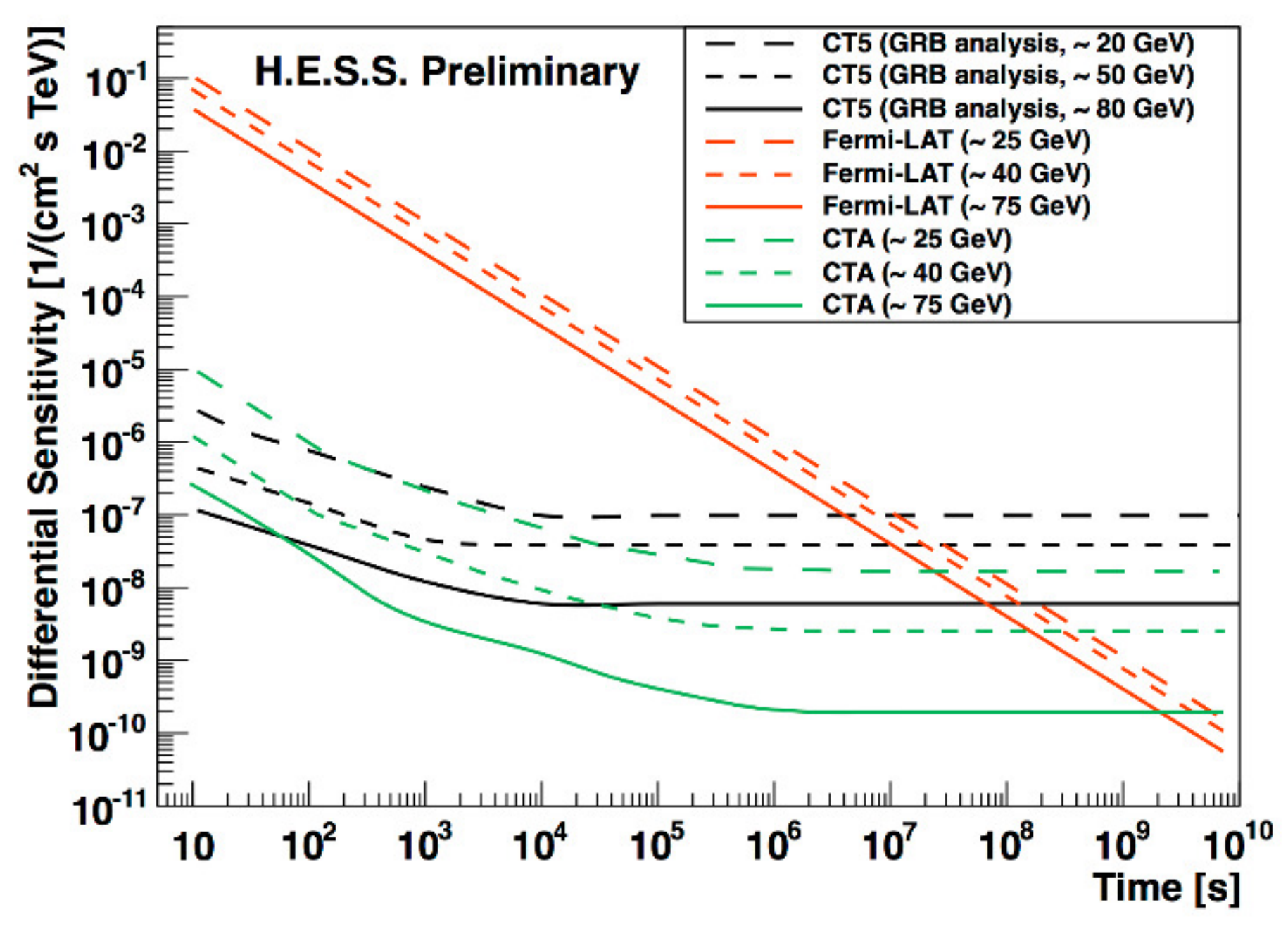}
  \includegraphics[width=.47\linewidth]{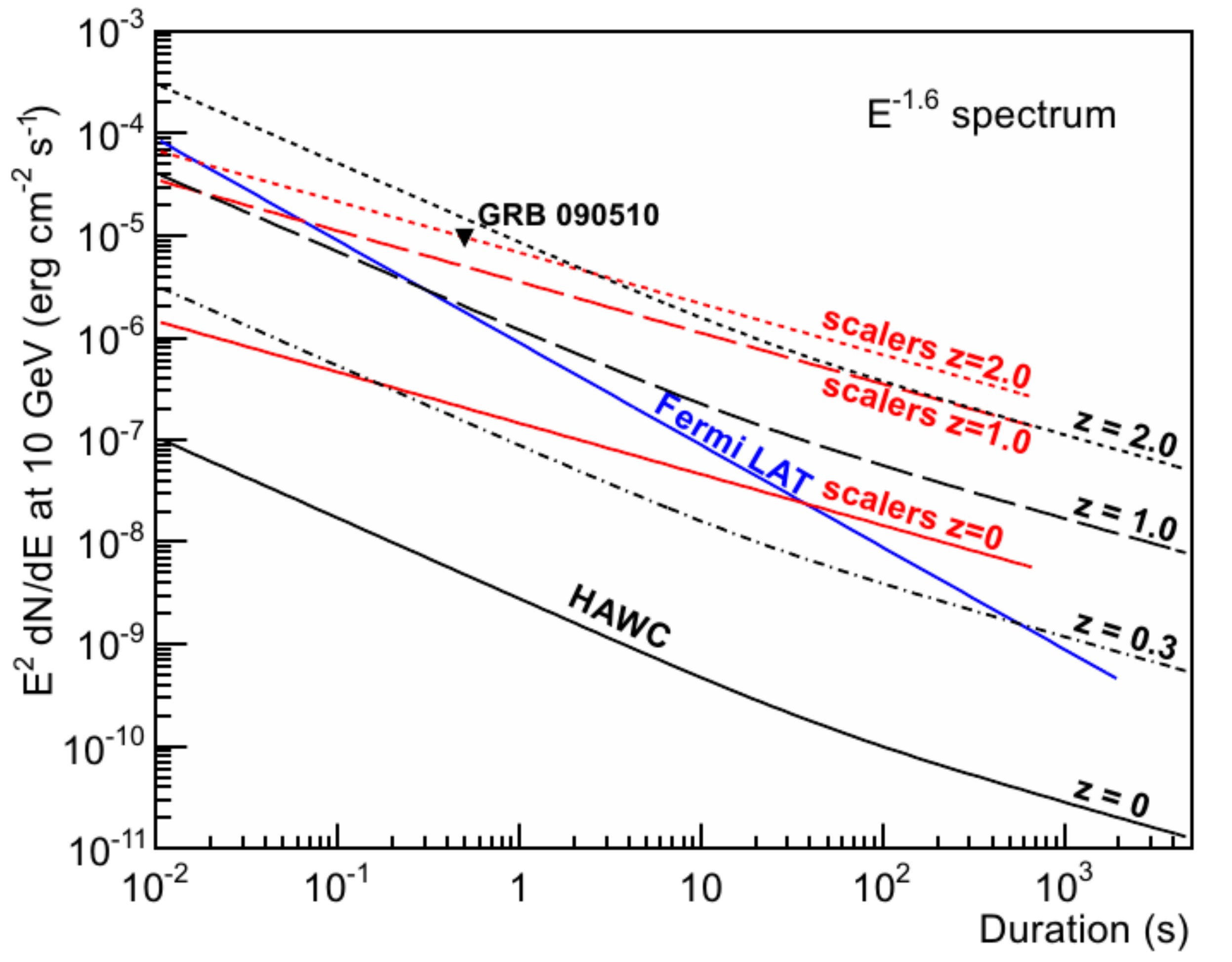}
  \caption{
    Sensitivity to transient sources of very high-energy telescopes (HESS CT5 and CTA, left panel; HAWC, right panel) as a function
    of the signal duration, for various energy thresholds (left panel) and redshifts (right panel), as compared to the \Fermi/LAT.
    CTA curves correspond to an analysis which has been optimized for GRB and pulsar studies~\cite{stegman2014}.
    HAWC curves (in erg\,cm$^{-2}$\,s$^{-1}=10^{-3}$\,W\,m$^{-2}$) correspond to a 20$^\circ$ zenith angle of
    the observed source, and account for the absorption of high-energy gamma rays by the Extra-galactic Background
    Light~\cite{hawc1}.
  }
  \label{fig:vhe}
\end{figure}

The observation by the \Fermi/LAT of several GRB photons with energies reaching 10--100\,GeV in the source frame
(figure~\ref{fig:emax}, right), as late as $\sim$1 day after the trigger in the case of GRB\,130427A (figure~\ref{fig:emax},
left), is an encouraging sign for GRB detections at very high energies with ground-based experiments, either with arrays of imaging
atmospheric Cherenkov telescopes (IACTs) such as
HESS\footnote{\tt{https://www.mpi-hd.mpg.de/hfm/HESS}},
MAGIC\footnote{\tt{https://magic.mpp.mpg.de}},
VERITAS\footnote{\tt{http://veritas.sao.arizona.edu}}
and the future CTA observatory\footnote{\tt{http://www.cta-observatory.org}}~\cite{actis},
or with synoptic detectors such as
HAWC\footnote{\tt{http://hawc.umd.edu}}
and LHAASO\footnote{\tt{http://english.ihep.cas.cn/ic/ip/LHAASO}}.
These experiments (see~\cite{cras2015-future} for more details) try to achieve low energy thresholds, ideally as
low as $\sim$10\,GeV in order to limit the absorption of high-energy gamma rays from distant sources by the Extra-galactic
Background Light.
Due to their huge effective area (of a few hectares, typically), they can accumulate large photon statistics and their
sensitivity to transient sources can surpass the \Fermi/LAT sensitivity beyond 10--100\,GeV (figure~\ref{fig:vhe}).\\

\begin{figure}[!t]
  \centering
  \includegraphics[width=\linewidth]{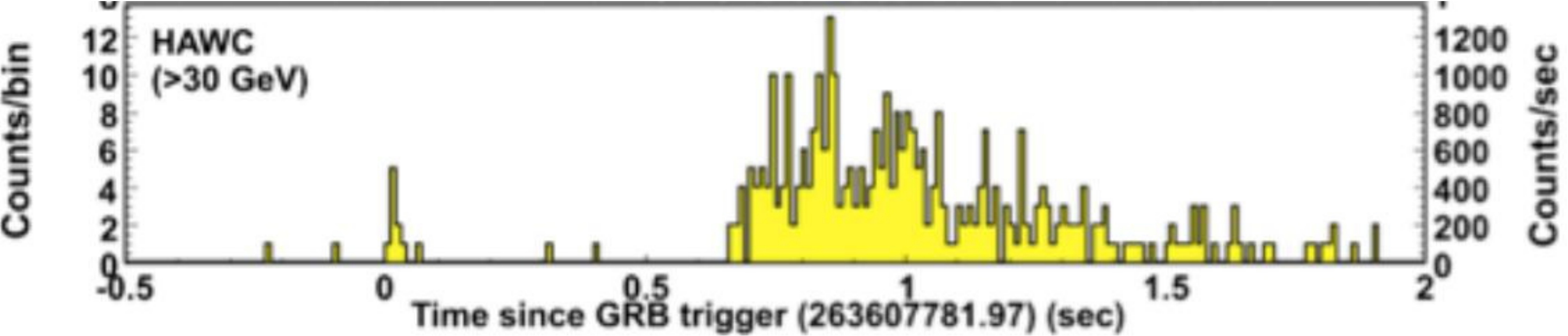}
  \caption{
    HAWC simulation of GRB\,090510 light curve above 30\,GeV, obtained by extrapolating the \Fermi/LAT spectrum to very high
    energies and including the effect of gamma-ray absorption by the Extra-galactic Background Light~\cite{tollefson13}.
  }
  \label{fig:hawc}
\end{figure}

Observations of GRBs with IACTs have always led to upper limits so far, on time scales ranging from a few tens of seconds to days.
Apart from the faintness of most GRBs at very high energies, the reasons for this lack of success include the relatively high
energy threshold ($\sim$100--200\,GeV) of past experiments, a low duty cycle ($\sim$10\%), as well as the combination of the IACT narrow
field of view (a few degrees) with the poor localization capabilities of the most productive space detectors (e.g., the CGRO/BATSE
and the \Fermi/GBM) and with the time needed to repoint and start follow-up observations.
Synoptic detectors have $\sim$1\,sr fields of view and a $\lesssim$100\% duty cycle, yet higher energy thresholds and lower
sensitivities than IACTs, due to their smaller power in rejecting the background events induced by charged cosmic rays.
As a result, the response time and sensitivity of IACTs and synoptic detectors match adequately the time scales and
brightness of long and short GRBs, respectively.
As an example, figure~\ref{fig:hawc} shows a simulation of the light curve that HAWC would have recorded above 30\,GeV from the short
GRB\,090510 if the experiment had been operational in 2009.\\

Extrapolating \Fermi/LAT GRB spectra to the very high-energy range is difficult, in particular it remains unclear whether the
extra power-law component observed in the spectra of several bright GRBs is a common property at GeV energies.
In addition, intrinsic spectral cutoffs similar to the case of GRB\,090926A are expected at 1--100\,GeV energies, and they are
strongly related to the value of the jet Lorentz factor whose distribution among GRBs is not precisely known.
For these reasons, current estimates of GRB detection rates at very high energies suffer from important uncertainties and amount
to $\approx1$ GRB per year depending on the considered experiment~\cite{hawc1,inoue2013,gilmore13,hawc2}.
During the coming years, a few but invaluable GRB detections are thus expected beyond 10--100\,GeV.
As VERITAS and HAWC upper limits on GRB\,130427A emission already suggest~\cite{130427a_veritas,130427a_hawc}, future detections at
very high energies will provide new constraints on GRB jet physics and useful information regarding their acceleration and
emission processes at the highest energies.

\section{Conclusions}
\label{sec:concl}

Since its launch in 2008, the \Fermi Gamma-ray Space Telescope has made important breakthroughs in the understanding of the GRB
phenomemon.
The combination of the \Fermi GBM and LAT instruments provided high quality data over seven decades in energy for a large sample
of GRBs, and it made GRB population studies possible at high energies.
It revealed that both short and long GRBs have relativistic outflows and share similar properties.
Their emission above 100\,MeV is delayed and temporally extended with respect to the emission detected at keV--MeV energies.
While the origin of the delayed onset remains unclear, the long-lived GeV emission is consistent with the afterglow
emission of a blast wave in adiabatic expansion, as confirmed by the stability of its spectrum which contrasts with the
spectral variability of the prompt keV--MeV emission.
However, high-energy observations of GRB\,130427A put severe constraints on forward shock synchrotron emission models.
Understanding the complex GRB spectral evolution during their prompt emission phase also requires new theoretical developments
and detections of more and brighter GRBs in the future, accompanied by more detailed time-resolved
spectroscopy, in order to pinpoint which high-energy processes dominate throughout the GRB.
In particular, the connection between the extra power-law component seen by the LAT at high energies in the prompt phase and
the long-lived GeV emission observed up to several (tens of) kilo-seconds, is of great importance in understanding the transition
from the internal shock phase to the early and late afterglow phases.\\

The \Fermi and \Swift observatories have provided (and still provide) a good characterization of GRB prompt and afterglow
emissions, respectively.
In the future, SVOM~\cite{svom} panchromatic observations from the near infra-red domain to MeV energies will bring new
spectro-temporal diagnosis during the entire period of activity of each GRB, from the prompt phase (and possible precursors) to the late
afterglow phase.
Scheduled for a launch in 2021, the SVOM mission will thus provide a broad view of the GRB phenomenon, which will be completed at higher
energies by new space-based gamma-ray observatories~\cite{cras2015-future} and, in a few cases, by observations
with ground-based facilities such as CTA and HAWC.



\end{document}